\documentclass{aa}

\usepackage{xcolor}
\usepackage{graphicx} 

\definecolor{lightgray}{rgb}{0.83, 0.83, 0.83}
\definecolor{lightblue}{rgb}{0.67, 0.84, 0.90}
\definecolor{lightgreen}{rgb}{0.56, 0.93, 0.56}

\usepackage{natbib}
\bibpunct{(}{)}{;}{a}{}{,} 

\usepackage{comment}
\usepackage{threeparttable}
\usepackage{makecell, tabularx}
\setcellgapes{5pt}

\usepackage[varg]{txfonts}
\usepackage{arydshln}
\usepackage{hyperref}
\usepackage{multirow}
\usepackage{color}
\definecolor{green}{rgb}{0.3,0.7,0.}
\definecolor{purple}{rgb}{0.77, 0.29, 0.55}

\newcommand\gva{{\sc Genec}}

\hypersetup{
    colorlinks=true,       
    linkcolor=blue,          
    citecolor=blue,        
    filecolor=blue,      
    urlcolor=blue           
}

\begin{document}

\title{Dark-Matter-Powered Population III Evolution: Lifetimes, Rotation, and Quasi-Homogeneity in massive Stars}
\titlerunning{Dark-Matter-Powered Population III Evolution}
\author{Anaïs Pauchet$^{1\ \star}$ and Devesh Nandal$^{2,3}$}
\authorrunning{Pauchet \& Nandal}

\institute{$^1$I. Physikalisches Institut, Universitat zu Köln, Zülpicher Str. 77, D-50937 Köln, Germany\\
$^2$Department of Astronomy, University of Virginia, 530 McCormick Rd, Charlottesville, VA 22904, USA\\$^3$Center for Astrophysics, Harvard and Smithsonian, 60 Garden St, Cambridge, MA 02138, USA\\
$^\star$ email: pauchet@ph1.uni-koeln.de}

\date{}

\abstract{
Population III stars supplied the first light and metals in the Universe, setting the pace of re-ionisation and early chemical enrichment.  
In dense haloes their evolution can be strongly influenced by the energy released when Weakly Interacting Massive Particles (WIMPs) annihilate inside the stellar core. We follow the evolution of a \(20\,M_\odot\) Population III model with the \textsc{genec} code, adding a full treatment of spin dependent WIMP capture and annihilation. Tracks are calculated for six halo densities from \(10^{8}\) to \(3\times10^{10}\,\mathrm{GeV\,cm^{-3}}\) and three initial rotation rates between zero and \(0.4\,v/v_{\mathrm{crit}}\). As soon as the capture product reaches \(\rho_\chi\sigma_{\mathrm{SD}}\simeq2\times10^{-28}\,\mathrm{GeV\,cm^{-1}}\), the dark-matter luminosity rivals hydrogen fusion, stretching the main-sequence lifetime from about ten million years to more than a gigayear.  
The extra time allows meridional circulation to smooth out differential rotation; a star that begins at \(0.4\,v/v_{\mathrm{crit}}\) finishes core hydrogen burning with near solid-body rotation and a helium core almost twice as massive as in the dark-matter-free case.  
Because the nuclear timescale is longer, chemically homogeneous evolution now sets in at only \(0.2\,v/v_{\mathrm{crit}}\), rather than the \(\gtrsim0.5\,v/v_{\mathrm{crit}}\) required without WIMPs.  
For a star with \(0.4\,v/v_{\mathrm{crit}}\), the surface hydrogen fraction drops to \(X\!\sim\!0.27\), helium rises to \(Y\!\sim\!0.73\), and primary \(^{14}\mathrm N\) increases by four orders of magnitude at He exhaustion. The star leaves the ZAMS sequence cooler, at \(T_{\mathrm{eff}}\approx50\) kK, and should display the strong N \textsc{iii} and He \textsc{ii} lines typical of a nitrogen-rich Wolf-Rayet analogue. Moderate rotation combined with plausible dark-matter densities can therefore drive primordial massive stars towards long-lived, quasi-homogeneous evolution with distinctive chemical and spectral signatures.  
Our tracks offer quantitative inputs for models of re-ionisation, for stellar archaeology, and for future attempts to constrain the microphysics of WIMPs through high-redshift observations.
}


\maketitle

\section{Introduction}

Understanding the nature of dark matter (DM) remains one of the most pressing challenges in astrophysics and cosmology. Observational evidence from galactic rotation curves, gravitational lensing, and cosmic microwave background anisotropies strongly suggests the existence of DM, which constitutes about 85\% of the matter content of the Universe \citep{PlanckColl-2018}. Despite its elusive nature, DM plays a crucial role in the formation and evolution of cosmic structures.

Among the various DM candidates, Weakly Interacting Massive Particles (WIMPs) are particularly compelling due to their natural emergence in extensions of the Standard Model of particle physics \citep{Jungman1996,Bertone-2005}. WIMPs interact through gravity and potentially via weak-scale interactions, allowing them to scatter off baryonic matter and annihilate with each other. In regions of high DM density, such as the early Universe or the centers of galaxies and stars, WIMP annihilation can release significant amounts of energy \citep{Bergstrom-1998}.

The first generation of stars, known as Population~III (Pop~III) stars, formed in DM halos at high redshifts ($z \sim 20$--30) \citep{Bromm-2004,Maeder2012, Liu2025}. These stars are thought to be massive and metal-free, playing a pivotal role in the reionization of the Universe, and the enrichment of the interstellar medium with heavy elements \cite{Liu2021}. The formation and evolution of Pop~III stars are influenced by the conditions of the early Universe, including the presence of DM.
In particular, the concept of ``dark stars'' (DS) has been proposed, where DM annihilation provides an additional energy source that can significantly alter stellar structure and evolution \citep{Spolyar2008,Freese2010, Ilie-2023}. These hypothetical stars form when the gravitational collapse of baryonic matter in early DM halos leads to the accumulation of both ordinary matter and DM particles. The annihilation of WIMPs within the star's core can counteract gravitational contraction, resulting in stars that are cooler, larger, and have extended lifetimes compared to conventional Pop~III stars \citep{Freese2008b,Ilie2012,Rindler-Daller2015,Tan2024, Nandal2025c}. The prolonged lifetimes and unique characteristics of dark stars may leave observable imprints, such as distinct spectral features or anomalous elemental abundances, which could be probed by future telescopes like the James Webb Space Telescope \citep{Zackrisson2010,freese-2016, Banik2019, Ilie2025}.

The interaction between DM and stars has been extensively studied over the past decades. \citet{gould_1987b} developed the theoretical framework for WIMP capture and annihilation within stars, laying the foundation for understanding how DM can influence stellar evolution. This framework was further refined to account for various stellar environments and DM properties \citep{Press1985,Faulkner1985}. Subsequent works have explored the impact of DM on stellar evolution in greater detail, including the potential formation of dark stars \citep{Freese2008a,Ilie2012,Scott-2009, Ilie-2021}. Studies have shown that DM annihilation can alter the temperature gradients within stars, affect nucleosynthesis processes, and modify the stellar evolution tracks on the Hertzsprung-Russell diagram \citep{taoso_dark_2008,Casanellas2011a, Casanellas2011b}. The effects of DM annihilation on Pop~III stars, considering both rotating and non-rotating models, have significant implications for our understanding of the first stars and the role of DM in stellar physics \citep{Yoon2008,Hirano2015,Stacy-2012}.

\citet{Tan2024} classified Pop~III stars into Pop~III.1 and Pop~III.2, based on their formation environments and the influence of radiation feedback. Pop~III.1 stars form in pristine environments without prior star formation, resulting in isolated, massive stars that can reach masses up to several hundred solar masses \citep{Hirano2014}. In contrast, Pop~III.2 stars form later, influenced by radiation and chemical feedback from earlier generations, leading to a broader mass spectrum and different formation pathways \citep{Greif2012}. The presence of DM can further complicate this classification, as DM annihilation may play a more pronounced role in the formation and evolution of Pop~III.1 stars due to the higher DM densities in early halos \citep{Natarajan2009}. Recent simulations have investigated the interplay between DM annihilation and stellar rotation, suggesting that rotation can enhance the mixing processes within stars and affect the transport of angular momentum and chemical elements \citep{Haemmerle2020,stacy2013}. Understanding these effects is crucial for predicting the observational signatures of the first stars and their contribution to cosmic reionization and early chemical enrichment \citep{Bromm2009,Wise2012, Gondolo-2022}.

In this work, we investigate the effects of DM annihilation on the evolution of rotating and non-rotating massive Pop~III stars. Using updated stellar evolution models, we analyze how DM capture and annihilation influence stellar structure, lifetime, and nucleosynthesis. By comparing models with and without rotation, we aim to elucidate the interplay between DM physics and stellar evolution in the context of primordial stars.

The paper is organized as follows: in Sect.~\ref{Sec:Methods}, we describe the methods and numerical setup used in our simulations. In Sect.~\ref{Sec:Results}, we present the results of our models and discuss the implications for Pop~III star evolution. Finally, in Sect.~\ref{sec:conclusions}, we summarize our findings and outline prospects for future work.

\section{Methods} \label{Sec:Methods}

In this work, we model the impact of dark matter (DM) annihilation on Pop~III stellar evolution by incorporating the capture and annihilation of weakly interacting massive particles (WIMPs) into the Geneva stellar evolution code (\gva). Our formalism builds on that of \citet{taoso_dark_2008} (hereafter T08) while reflecting improvements in the underlying physics and numerical techniques \citep{Ekstrom2012,Nandal2024b, Nandal2025b}. In the following, we detail the methodology and present the relevant equations, clarifying all assumptions and definitions.

The evolution of the total number of WIMPs in the star, \(N_\chi\), is governed by
\begin{equation} \label{eq:dNdt}
    \frac{dN_\chi}{dt} = C_c + C_s\,N_\chi - A\,N_\chi^2 - E\,N_\chi,
\end{equation}
where:
\begin{itemize}
    \item \(C_c\) is the capture rate via WIMP--baryon scattering.
    \item \(C_s\) is the self-capture rate due to WIMP--WIMP interactions.
    \item \(A\) is the annihilation coefficient, representing the rate at which pairs of WIMPs annihilate.
    \item \(E\) denotes the evaporation rate.
\end{itemize}
For the massive stars studied here, both self-capture and evaporation are negligible; hence, these terms are set to zero in our models. The total number of WIMPs is defined as
\[
N_\chi = \int_0^{R_*} n_\chi(r)\, dV,
\]
with \(n_\chi(r)\) being the local WIMP number density and \(R_*\) the stellar radius.

We adopt a Gaussian profile to describe the spatial distribution of WIMPs within the star:
\begin{equation}
    n_\chi(r) = n_0 \, e^{-r^2/r_\chi^2},
\end{equation}
where the characteristic radius,
\begin{equation}
    r_\chi = \sqrt{\frac{3k T_c}{2\pi G \rho_c m_\chi}},
\end{equation}
marks the region where WIMP--baryon interactions are most effective; here \(T_c\) and \(\rho_c\) are the central temperature and density, \(m_\chi\) is the WIMP mass, \(k\) is Boltzmann's constant, and \(G\) is the gravitational constant.

\subsection{Capture Rate} \label{Sec:Capture}

WIMPs entering the star lose kinetic energy via scattering off baryons until they become gravitationally bound. The local differential capture rate is expressed as
\begin{multline}\label{eq:dCdV}
    \frac{dC_c(r)}{dV} = \left(\frac{6}{\pi}\right)^{1/2} \sigma_{\chi,N} \frac{\rho_i(r)}{M_i} \frac{\rho_\chi}{m_\chi} \frac{v^2(r)}{\bar{v}^2} \frac{\bar{v}}{2\eta A^2} \\
    \times  \left\{\left(A_+A_- - \frac{1}{2}\right) \left[\chi (-\eta,\eta) - \chi(A_-, A_+)\right] \right. \\
    \left. + \frac{1}{2}A_+ e^{-A_-^2} - \frac{1}{2}A_- e^{-A^2_+} - \frac{1}{2} \eta e^{-\eta^2} \right\}.
\end{multline}
Here, the variables are defined as follows:
\begin{align*}
    A^2 &= \frac{3\,v^2(r)\,\mu}{2\bar{v}^2\,\mu_-^2}, \qquad & A_\pm &= A \pm \eta, \\
    \eta^2 &= \frac{3\,v_*^2}{2\bar{v}^2}, \qquad & \chi(a,b) &= \frac{\sqrt{\pi}}{2}\left[\operatorname{Erf}(b)-\operatorname{Erf}(a)\right], \\
    \mu_i &= \frac{m_\chi}{M_i}, \qquad & \mu_- &= \frac{\mu_i-1}{2}.
\end{align*}
In these expressions, \(\rho_i(r)\) is the density profile of element \(i\) (with atomic mass \(M_i\)), \(\sigma_{\chi,N}\) is the WIMP--baryon scattering cross section (which can be spin-dependent or spin-independent), and \(\rho_\chi\) and \(\bar{v}\) are, respectively, the ambient WIMP density and velocity dispersion. The stellar velocity \(v_*\) is assumed to be equal to \(\bar{v}\). The escape velocity \(v(r)\) is computed from
\begin{equation}
    v^2(r) = \int_{\infty}^{r} \frac{GM(r')}{r'^2} dr',
\end{equation}
where \(M(r')\) is the mass enclosed within radius \(r'\).

The total capture rate is obtained by integrating Eq.~\eqref{eq:dCdV} over the entire stellar volume:
\begin{equation}\label{eq:C}
    C_c = 4\pi \int_{0}^{R_*} r^2 \, \frac{dC_c(r)}{dV}\, dr.
\end{equation}

\subsection{Annihilation Rate} \label{Annihilation}

Assuming that WIMPs are their own antiparticles \citep{Jungman1996,Aprile_2017}, they can annihilate in the dense stellar core. The local energy generation rate per unit mass due to WIMP annihilation is given by
\begin{equation}\label{eq:eps_a}
    \epsilon_a(r) = \frac{1}{2}\langle \sigma v \rangle\, m_\chi c^2 \frac{n_\chi^2(r)}{\rho(r)},
\end{equation}
where \(\langle\sigma v\rangle\) is the thermally averaged annihilation cross section, \(c\) is the speed of light, and \(\rho(r)\) is the local baryon density. The integrated annihilation coefficient is then defined as
\begin{equation}
    A = 4\pi \int_0^{R_*} \epsilon_a(r)\, \rho(r)\, r^2\, dr.
\end{equation}

We assumed that the totality of the annihilation energy is processed by the star. Hence 
\begin{equation}
L_\chi = m_\chi C_c   
\end{equation}
the net luminosity deposited in the star. Works by \cite{Freese2010, Rindler-Daller2015, Ilie-2021, Ilie2025} consider that a third of the energy is lost by neutrinos escaping the star.

\subsection{Normalization and Equilibrium}

A key unknown in our formulation is the normalization constant \(n_0\) of the WIMP density profile. To determine \(n_0\), we assume that after a characteristic timescale
\[
t_\chi = \sqrt{\frac{1}{C_c \langle \sigma v \rangle} \, \pi^{-3/2}\, r_\chi^{-3}},
\]
the system attains equilibrium such that the rate of WIMP annihilation balances the capture rate. Under the simplifying assumption that self-capture and evaporation are negligible, the equilibrium condition can be approximated by
\[
2A\,N_\chi^2 = C_c.
\]
Once \(C_c\) is computed via Eq.~\eqref{eq:C}, \(n_0\) is adjusted in GENEC so that the energy injection from WIMP annihilation, as given by Eq.~\eqref{eq:eps_a}, satisfies this equilibrium condition. Although this approach involves approximations regarding the equilibrium timescale \(t_\chi\), the resulting uncertainties remain within acceptable limits for our analysis.

This methodology self-consistently integrates DM capture and annihilation into the stellar energy budget, allowing us to investigate how WIMP annihilation modifies stellar structure and evolution.

\begin{table*}[h!]
    \centering
    \caption{Name of the simulations and their specific parameters, the WIMPs density surrounding the star, the star initial rotation velocity, the SD cross section parameter used and the helium core mass, the velocity and the lifetime of the star at the end of the MS. 3 aligned crosses describe a simulation that did not reached the MS due to numerical problems. The names comporting \texttt{\_new} describe models with updated $\sigma_{\rm SD}$ value and corresponding $\rho_\chi$ to keep a similar behavior.}
    {
    \begin{tabular}{|c|cccccc|}
    \hline
    Name & $\rho_\chi$ $\left[\text{GeV}\cdot\text{cm}^{-3}\right]$ & $v/v_{crit}$ $\left[\%\right]$ & $\sigma_{SD}$ $\left[cm^{-2}\right]$ &  $M_{\text{He}}$ $\left[\text{M}_\odot\right]$ & $v_{\text{eq}}$ $\left[\text{km}\cdot \text{s}^{-1}\right]$  & Lifetime $\left[\text{Myr}\right]$ \\  \hline \hline
    \texttt{wtt}          &          0           & 0  & $10^{-38}$ & 5.056 &  0 & 9.64 \\
    \texttt{1d30} & $1.0\times 10^{8} $  & 0  & $10^{-38}$ & N/A   &  0 & 9.74\\
    \texttt{1d29}   & $1.0\times 10^{9} $  & 0  & $10^{-38}$ & N/A   &  0 & 10.77\\
    \texttt{3d29}& $3.0\times 10^{9} $  & 0  & $10^{-38}$ & N/A   &  0 & 13.89\\
    \texttt{63d29}& $6.3\times 10^{9} $  & 0  & $10^{-38}$ & N/A   &  0 & 23.50\\ 
    \texttt{1d28}& $1.0\times 10^{10} $ & 0  & $10^{-38}$ & 5.189 &  0 & 50.91\\
    \texttt{13d28} & $1.3\times 10^{10} $ & 0  & $10^{-38}$ & N/A   &  0 & 109.93\\
    \texttt{16d28}& $1.6\times 10^{10} $ & 0  & $10^{-38}$ & N/A   &  0 & 251.80\\ 
    \texttt{2d28}& $2.0\times 10^{10} $ & 0  & $10^{-38}$ & N/A   &  0 & 716.92\\
    \texttt{3d28}& $3.0\times 10^{10} $ & 0  & $10^{-38}$ & 5.360 &  0 & 8439.95\\ \hline
    
    \texttt{wtt\_0.2v}&          0           & 20 & $10^{-38}$ & 5.291  & 208 & 10.89\\
    \texttt{1d30\_0.2v}& $1.0\times 10^{8} $  & 20 & $10^{-38}$ & 5.306  & 209 & 11.03\\
    \texttt{1d29\_0.2v}& $1.0\times 10^{9} $  & 20 & $10^{-38}$ & 5.396  & 216 & 12.54\\
    \texttt{3d29\_0.2v}& $3.0\times 10^{9} $  & 20 & $10^{-38}$ & 5.626  & 231 & 17.38\\
    \texttt{63d29\_0.2v}& $6.3\times 10^{9} $  & 20 & $10^{-38}$ & 6.260  & 264 & 34.07\\
    \texttt{1d28\_0.2v}& $1.0\times 10^{10} $ & 20 & $10^{-38}$ & 7.324  & 319 & 84.16\\
    \texttt{13d29\_0.2v}& $1.3\times 10^{10} $ & 20 & $10^{-38}$ & 8.250  & 368 & 183.30\\
    \texttt{16d29\_0.2v}& $1.6\times 10^{10} $ & 20 & $10^{-38}$ & 9.068  & 421 & 373.71\\
    \texttt{2d28\_0.2v}& $2.0\times 10^{10} $ & 20 & $10^{-38}$ & 9.983  & 489 & 791.30\\
    \texttt{3d28\_0.2v}& $3.0\times 10^{10} $ & 20 & $10^{-38}$ &  X     & X   & X\\ \hdashline
    \texttt{wtt\_0.4v}&          0           & 40 & $10^{-38}$ & 6.795  & 546 & 14.62\\
    \texttt{63d29\_0.4v}& $6.3\times 10^{9} $  & 40 & $10^{-38}$ & 10.375 & 860 & 48.38\\
    \texttt{1d28\_0.4v}& $1.0\times 10^{10} $ & 40 & $10^{-38}$ & 13.281 & 963 & 91.96\\ \hdashline
    \texttt{1d28\_0.6v}& $1.0\times 10^{10} $ & 60 & $10^{-38}$ &   X    &  X  & X \\
     \hline
    
    \texttt{1d30\_new}& $1.0\times 10^{10} $ & 0  & $10^{-40}$ & 5.056 & 0  & 9.74\\ 
    \texttt{1d29\_new}& $1.0\times 10^{11} $ & 0  & $10^{-40}$ & 5.059 & 0  & 10.77\\
    \texttt{3d29\_new}& $3.0\times 10^{11} $ & 0  & $10^{-40}$ & 5.071 & 0  & 13.89\\
    \texttt{63d29\_new}& $6.3\times 10^{11} $ & 0  & $10^{-40}$ & 5.127 & 0  & 23.50\\ 
    \texttt{1d28\_new}& $1.0\times 10^{12} $ & 0  & $10^{-40}$ & 5.189 & 0  & 50.91\\
    \texttt{13d28\_new}& $1.3\times 10^{12} $ & 0  & $10^{-40}$ & 5.244 & 0  & 109.94\\ 
    \texttt{16d28\_new}& $1.6\times 10^{12} $ & 0  & $10^{-40}$ & 5.275 & 0  & 251.83\\ 
    \texttt{2d28\_new}& $2.0\times 10^{12} $ & 0  & $10^{-40}$ & 5.315 & 0  & 717.01\\ 
    \texttt{3d28\_new}& $3.0\times 10^{12} $ & 0  & $10^{-40}$ & 5.360 & 0  & 8445.11\\ \hline

    \end{tabular}
    }
    \label{tab:all_data}
\end{table*}

\section{Results}\label{Sec:Results}
 We first summarize the suite of 20M$_\odot$ PopIII star models in Table~\ref{tab:all_data}. Columns1–4 list the model name and key initial parameters: ambient WIMP density $\rho_\chi$, initial rotation (as a percentage of the critical velocity $v_{\rm crit}$), and the spin-dependent WIMP scattering cross-section $\sigma_{\rm SD}$. Columns5–7 give selected results at the end of core hydrogen burning: the helium core mass $M_{\rm He}$, the equatorial surface velocity $v_{\rm eq}$, and the stellar lifetime. Non-rotating models ($v/v_{\rm crit}=0$\%) are listed in the upper part of the table, followed by rotating cases (20\%, 40\%, and 60\% of $v_{\rm crit}$). In cases where the main sequence (MS) evolution did not fully converge (defined here by central hydrogen mass fraction $X_c$ reaching $10^{-3}$), the table entries are marked with “X”. As seen in Table~\ref{tab:all_data}, increasing the ambient WIMP density (for fixed $\sigma_{\rm SD}$) dramatically extends the MS lifetime of a 20~M$\odot$ star. For example, in the non-rotating sequence at $\sigma_{\rm SD}=10^{-38}$~cm$^2$, the lifetime grows from $\sim9.6$~Myr with no DM up to $\sim8.44\times10^3$~Myr at the highest DM density ($\rho_\chi = 3\times10^{10}$GeVcm$^{-3}$). The helium core mass $M_{\rm He}$ remains around $5$~M$_\odot$ for those non-rotating models that complete H-burning, indicating that despite the vastly different MS durations, they synthesize similar helium cores. In contrast, the rotating models (which will be discussed in Sect.3.2) develop larger $M_{\rm He}$ and higher $v_{\rm eq}$ by MS termination due to rotational mixing. A second set of non-rotating models (Table\ref{tab:all_data}, last rows) explores a lower cross-section $\sigma_{\rm SD}=10^{-40}$~cm$^2$ with proportionally increased $\rho_\chi$: these yield evolution outcomes very similar to the $\sigma_{\rm SD}=10^{-38}$ cases, reflecting the primary dependence on the product $\rho_\chi \sigma_{\rm SD}$.

\subsection{Non-rotating models}\label{sec:nonrotating}

\subsubsection{Hertzsprung-Russell evolution}
\begin{figure}
    \centering
    \includegraphics[width=1\linewidth]{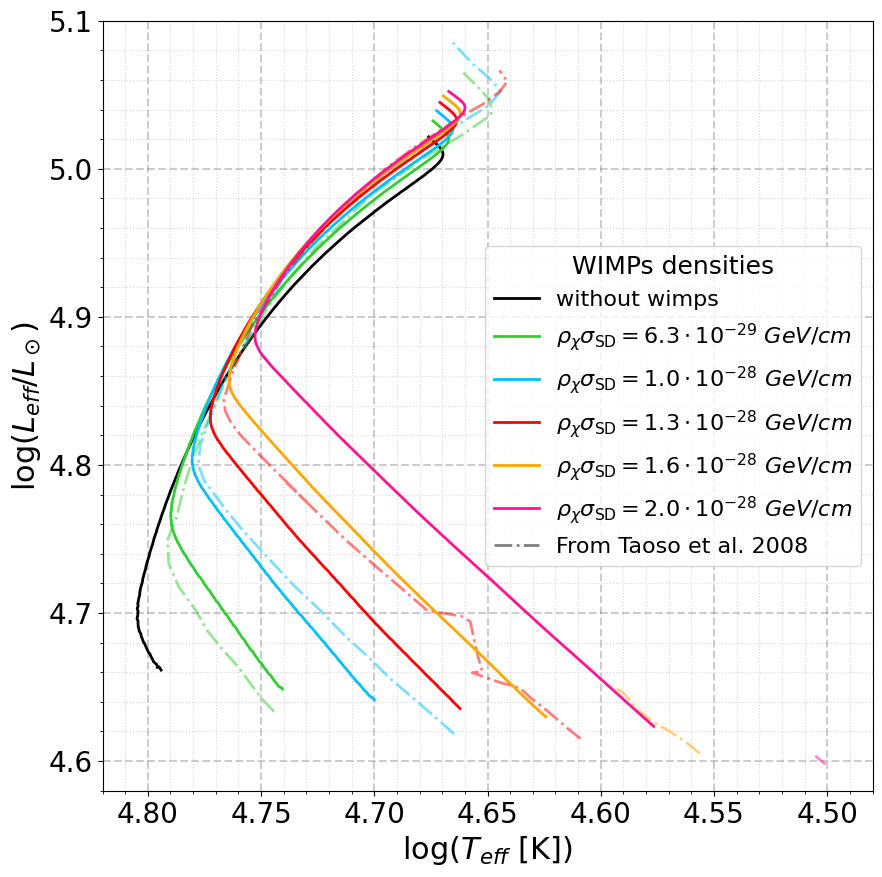}
    \caption{Hertzsprung–Russell diagram from this work (solid lines) and T08 (dashed lines) for a static 20 M$_\odot$ Pop III star surrounded by different WIMPs densities shown with specific colors.}
    \label{fig:HRD_norot}
\end{figure}

Figure~\ref{fig:HRD_norot} illustrates the evolutionary tracks of the non-rotating models in the Hertzsprung–Russell diagram (HRD). Increasing the ambient WIMP density shifts the Zero-Age Main Sequence (ZAMS) to markedly lower effective temperatures and slightly lower luminosities. Physically, the additional energy injection from WIMP annihilation in the stellar core inflates the star, increasing internal radiative pressure and yielding a larger radius (cooler $T_{\rm eff}$) at the ZAMS. Throughout the early-MS, the energy output is dominated by WIMP annihilation rather than nuclear fusion. Consequently, the star undergoes only minimal hydrogen burning during this phase, and its track in the HRD remains near the ZAMS locus for an extended period. In the HRD (Fig.~\ref{fig:HRD_norot}), this behavior is seen as an initial clustering of the tracks at relatively cool $T_{\rm eff}$ for high $\rho_\chi \sigma_{\rm SD}$ models. As evolution proceeds, however, the DM energy input gradually becomes less dominant. Once a significant amount of hydrogen is converted into helium in the core, the WIMP capture rate drops (since we assume purely spin-dependent capture; helium, with no nuclear spin, is ineffective at capturing WIMPs). Thereafter, nuclear energy production steadily increases and the star resumes a more conventional MS evolution. Accordingly, the HRD tracks for different $\rho_\chi \sigma_{\rm SD}$ eventually bend back toward higher $T_{\rm eff}$, converging with the track of a star without DM by the end of the MS. Aside from an initial offset in ZAMS position, the overall morphology of our HRD tracks with DM is similar to that found by \citet{taoso_dark_2008} (dashed curves in Figure~\ref{fig:HRD_norot}). Small quantitative differences (e.g. a slight shift in our ZAMS location) can be attributed to updates in the GENEC code since T08. Importantly, once WIMP annihilation becomes subdominant late in the MS, all tracks, with and without DM, converge to the same Hertzsprung–Russell endpoint.

\subsubsection{Main–sequence lifetimes}\label{sec:nonrot_lifetimes}

The non-rotating series in Table~\ref{tab:all_data} quantifies the dramatic lifetime extension produced by WIMP heating.  
The reference model without dark matter completes core–hydrogen burning in $9.6$\,Myr, whereas a mild enhancement  
$\rho_\chi = 6.3\times10^{9}$\,GeV\,cm$^{-3}$ (\texttt{63d9}, $\rho_\chi\sigma_{\rm SD}=6.3\times10^{-29}$\,GeV\,cm$^{-1}$) already doubles the lifetime to $23.5$\,Myr.  
Further increases give $50.9$\,Myr at $1.0\times10^{10}$ (\texttt{1d10}),  
$7.2\times10^{2}$\,Myr at $2.0\times10^{10}$ (\texttt{2d10}),  
and $8.4$\,Gyr at $3.0\times10^{10}$ (\texttt{3d10})—a factor of $10^{3}$ relative to the standard Pop\,III case. The alternate sequence with $\sigma_{\rm SD}=10^{-40}$\,cm$^{2}$ reproduces the same lifetimes when $\rho_\chi$ is increased by the same factor, demonstrating that the controlling parameter is the product $\rho_\chi\sigma_{\rm SD}$.  

The reference model without dark matter completes core–hydrogen burning in $9.6$\,Myr, whereas a mild enhancement $\rho_\chi\sigma_{\rm SD}=6.3\times10^{-29}$\,GeV\,cm$^{-1}$ (\texttt{63d29}) already doubles the lifetime to $23.5$\,Myr.  
Further increases give $50.9$\,Myr at $1.0\times10^{-28}$ (\texttt{1d28}),  
$7.2\times10^{2}$\,Myr at $2.0\times10^{-28}$ (\texttt{2d28}),  
and $8.4$\,Gyr at $3.0\times10^{28}$ (\texttt{3d28}) —a factor of $10^{3}$ relative to the standard Pop\,III case. The alternate sequence with $\sigma_{\rm SD}=10^{-40}$\,cm$^{2}$ reproduces the same lifetimes when $\rho_\chi$ is increased by the same factor, demonstrating that the controlling parameter is the product $\rho_\chi\sigma_{\rm SD}$.

During the early MS, energy injection from WIMP annihilation maintains a comparatively cool, low-density core; hydrogen fusion therefore proceeds slowly, and the stellar structure evolves on an extended nuclear timescale.  As hydrogen is converted to helium, spin-dependent capture weakens, nuclear burning gradually overtakes the dark-matter contribution, and the model ultimately resumes a standard terminal-age main-sequence path.  For $\rho_\chi\sigma_{\rm SD}\gtrsim10^{-28}$\,GeV\,cm$^{-1}$ the main-sequence duration exceeds a gigayear, implying that such WIMP-supported Pop\,III stars could persist to low redshift if their dark-matter environment remains undisturbed.

\begin{figure*}
    \centering
    \includegraphics[width=0.48\linewidth]{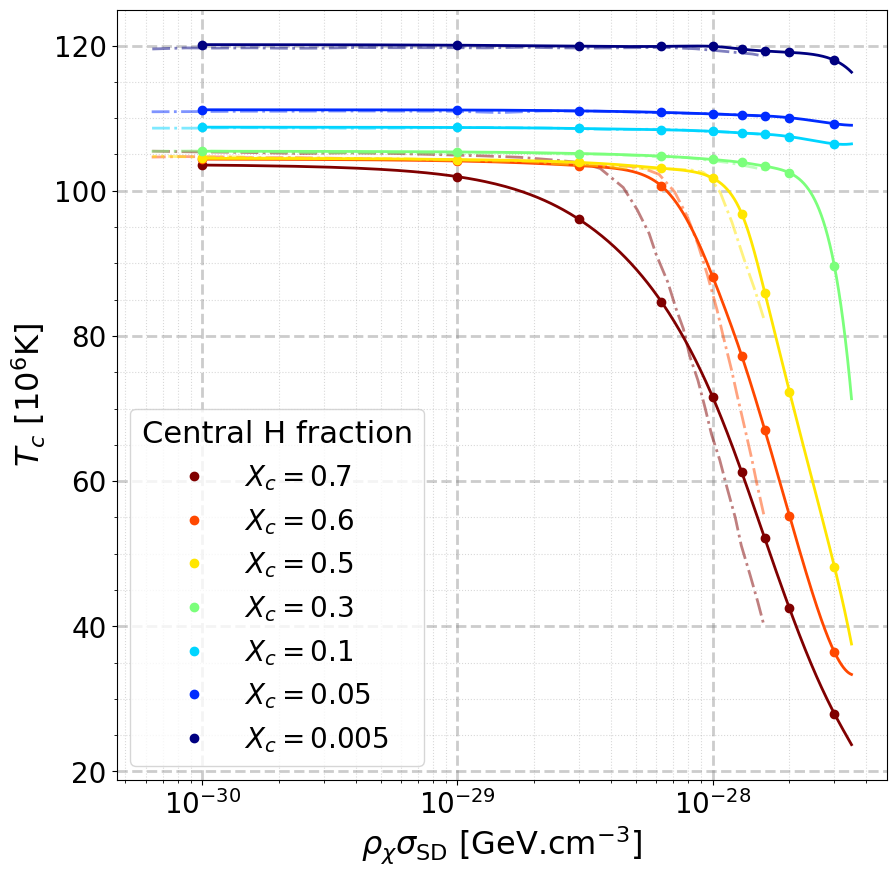}
    \includegraphics[width=0.48\linewidth]{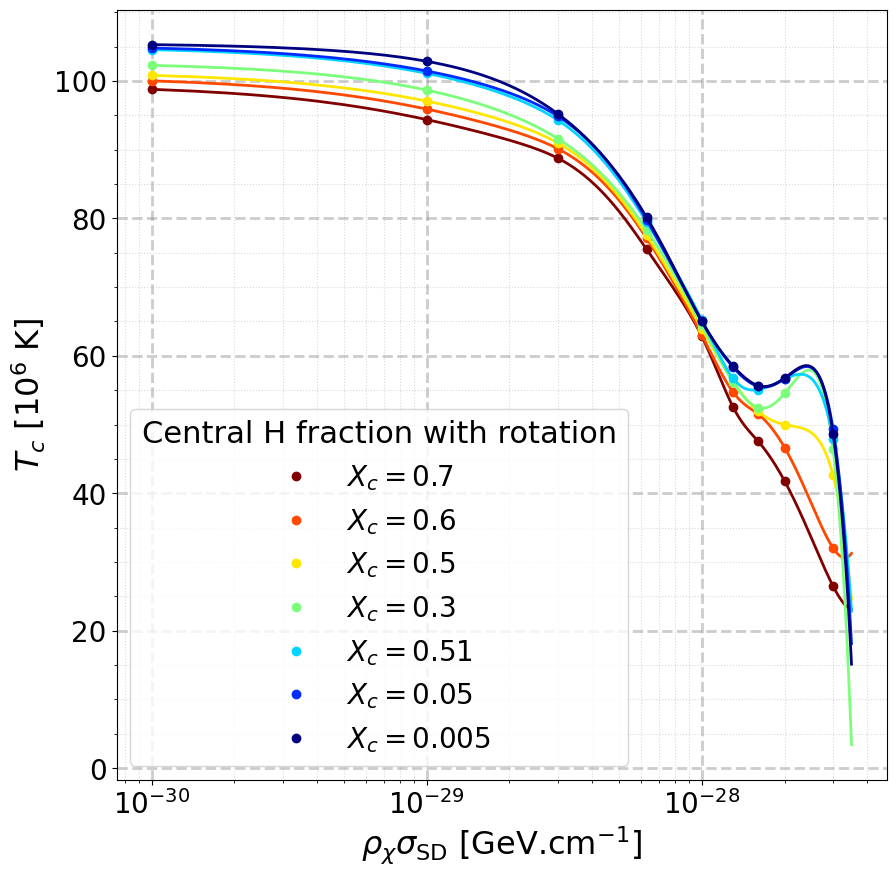}
    \caption{Central temperature as a function of WIMP density for a 20\,M\(_\odot\) Pop III star without rotation \textit{left}, and with a rotation 20\% of $v_{\rm crit}$ \textit{right}. Curves corresponding to different evolutionary stages (defined by the central hydrogen abundance) are plotted with spline interpolation; the dashed curves are taken from T08.}
    \label{fig:core_vs_rhochi}
\end{figure*}

\subsubsection{Internal structure}\label{sec:nonrot_structure}

Figure~\ref{fig:core_vs_rhochi} presents the response of the stellar core to increasing WIMP capture for seven central hydrogen mass fraction during the MS phase (from $X_{c}=0.70$ to $0.005$).\\
The left-hand panel shows the central temperature $T_{c}$ for a non rotating star, while the right-hand panel displays a star rotating with initial velocity 20\% of $v_{\rm}$.  
Both models are plotted against the product $\rho_\chi\sigma_{\rm SD}$, which parametrizes the WIMP capture rate.  

At ZAMS ($X_{c}=0.70$) the model without dark matter occupies $(T_{c},\rho_{c})\simeq(8.0\times10^{7}\,{\rm K},1.2\times10^{2}\,{\rm g\,cm^{-3}})$. A tenfold increase in $\rho_\chi\sigma_{\rm SD}$ lowers each quantity by roughly one quarter, and at $3\times10^{-28}$\,GeV\,cm$^{-1}$ both are reduced by nearly a factor of two, indicating that instead of hydrogen fusion, WIMP annihilation provides the dominant luminosity. The mid–main‐sequence curves ($X_{c}=0.40$) illustrate how this stabilising influence persists. In the standard model the core has already contracted and heated to $(9.5\times10^{7}\,{\rm K},1.5\times10^{2}\,{\rm g\,cm^{-3}})$, whereas DM–rich cores remain within $\sim10$\,\% of their ZAMS state, confirming that the additional energy source largely suppresses Kelvin–Helmholtz contraction during the first few $10^{7}$\,yr.  

Once hydrogen is exhausted, spin–dependent capture ceases and the tracks for all $\rho_\chi\sigma_{\rm SD}$ converge to a narrow locus around $T_{c}\simeq1.0$–$1.1\times10^{8}$\,K and $\rho_{c}\simeq1.6$–$1.8\times10^{2}$\,g\,cm$^{-3}$.  
Thus, despite large differences during most of the extended main sequence, each model ultimately attains the core conditions required for helium ignition after WIMP heating subsides.  In effect, dark‐matter energy support delays—but does not prevent—the standard contraction–heating sequence; the star therefore resumes canonical Pop\,III evolution once its hydrogen reservoir no longer sustains efficient WIMP capture.

\subsection{Rotating models}\label{sec:rotating}

Rotating Pop\,III models were evolved with GENEC’s standard meridional-circulation and shear prescription.  
We first compare stars that begin at 20\,\%\,$v_{\rm crit}$\ with the non-rotating sequence, then increase the spin to 40 and 60 \%\,$v_{\rm crit}$.  
Unless stated otherwise the spin-dependent cross-section is $\sigma_{\rm SD}=10^{-38}$ cm$^{2}$.  
The companion grid with $\sigma_{\rm SD}=10^{-40}$ cm$^{2}$ reproduces the same trends after $\rho_\chi$ is raised by the same factor and is noted briefly in Sect.~\ref{sec:nonrot_lifetimes}.

\subsubsection{Moderate rotation (20 \%\,$v_{\rm crit}$) vs.\ no rotation}\label{sec:rot20}

\begin{figure}
  \centering
  \includegraphics[width=\linewidth]{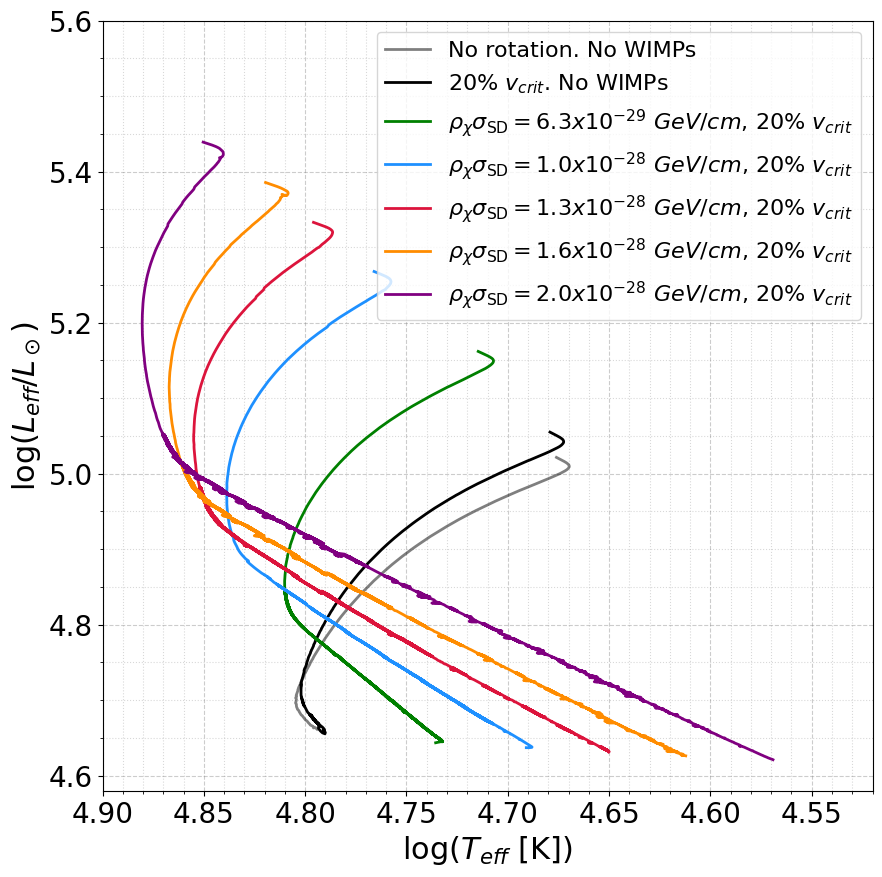}
  \caption{Similar plot as \ref{fig:HRD_norot} with an initial rotation velocity of 20 \%\, $v_{\rm crit}$. The grey curve represents the non-rotating tracks.}
  \label{fig:HRD_rot-all}
\end{figure}

The right-hand side of Figure~\ref{fig:core_vs_rhochi} shows the central temperature as a function of the product $\rho_\chi \sigma_{\rm crit}$ for a Pop III rotating star 20\% of $v_{\rm crit}$. In opposition to the static case the temperature stays lower at every step during the MS. With the mixing the energy produce by WIMPs annihilation is transported outwards, keeping the core cooler.

Figure~\ref{fig:HRD_rot-all} shows that rotation leaves the ZAMS almost unchanged but shifts the subsequent dark-star phase upward in luminosity.  
\texttt{wtt\_0.2v} is $\simeq0.05$ dex brighter and $5.0\times10^{2}$ K hotter, from the static model, at hydrogen exhaustion.  
With \texttt{1d28\_0.2v} the luminosity offset rises to 0.20 dex, while the temperature excess falls to $1.1\times10^{4}$ K because centrifugal support partly counteracts DM-driven contraction.\\
The higher $L$ reflects a larger convective core: rotational mixing carries fresh H inward and He outward, raising the nuclear luminosity that supplements WIMP heating.  
Consequently the envelope expands less than in the non-rotating case, keeping the surface hotter for a given $\rho_\chi$.  

Quantitatively, the helium-core mass at MS exit grows from 5.06\,M$_\odot$ (\texttt{wtt}) to 5.29\,M$_\odot$ for \texttt{wtt\_0.2v}, and from 5.19\,M$_\odot$ to 7.32\,M$_\odot$ for \texttt{1d28\_new} and \texttt{1d28\_0.2v} respectively.  
The factor-of-1.4 increase sets the stage for still greater differences at higher spin.

\subsubsection{Faster rotation: 40–60 \%\,$v_{\rm crit}$} \label{sec:rot4060}

\begin{figure}
  \centering
  \includegraphics[width=\linewidth]{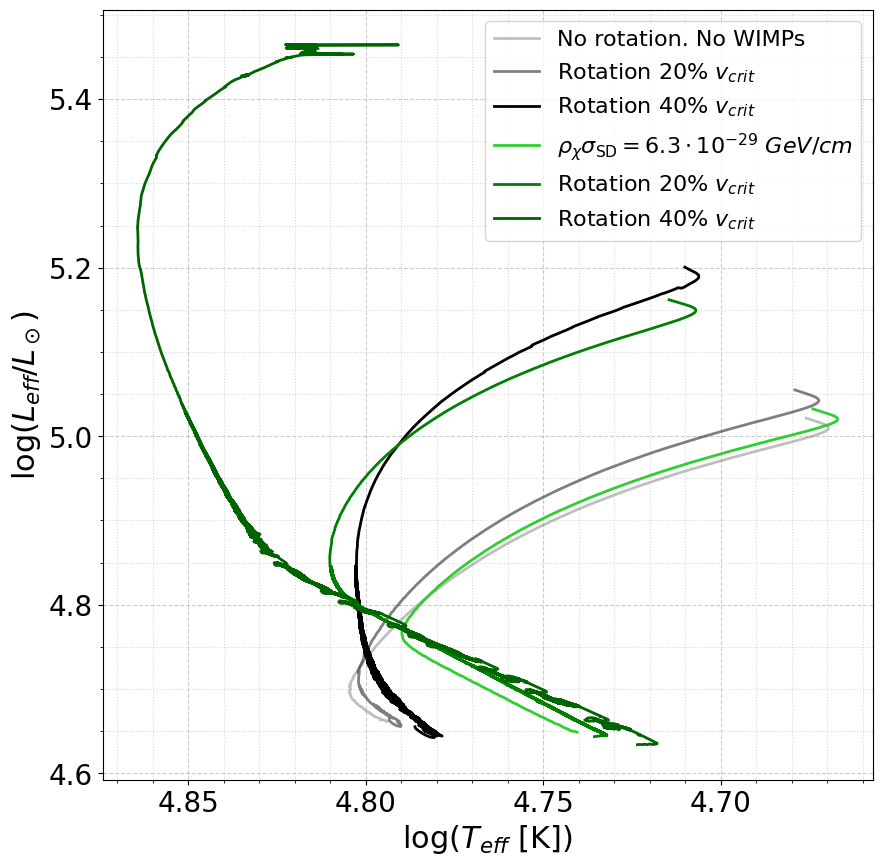}
  \caption{HRD of a Pop III star of 20 M$_\odot$, from the ZAMS to the end of the MS, for a WIMPs density of $\rho_\chi \sigma_\text{SD} = 6.3\cdot 10^{-29}$ $\text{GeV}\cdot\text{cm}^{-1}$, for an initial veloicity at 20\% and 40\% of the critical velocity of the star}
  \label{fig:HRD_rot-63d9}
\end{figure}

The 40\% $v_{\rm crit}$ (\texttt{63d29\_0.4v}) track in Fig.~\ref{fig:HRD_rot-63d9} is $\simeq 0.3$\,dex more luminous and $\sim1.6\times10^{4}$ K hotter than the 20\% (\texttt{63d29\_0.2v}) track at the terminal‐age MS ($X_{c}\simeq10^{-3}$).  
The luminosity rise follows directly from the larger helium core: $M_{\rm He}=13.28$ M$_\odot$ versus 7.32 M$_\odot$, a factor 1.8 increase that boosts the nuclear energy release once WIMP heating subsides.  
Centrifugal support lowers the effective gravity at the stellar surface by $\approx12$ \% (evaluated at the equator), reducing the photospheric temperature despite the higher $L$. Rotationally induced meridional circulation also transports entropy outward, inflating the envelope by $\sim20$ \% in radius relative to the 20 \% model. The net result is a star that is brighter but slightly cooler, sliding upward and rightward in the HRD. A higher DM supply ($\rho_\chi\sigma_{\rm SD}=1.0\times10^{-28}$ GeV cm$^{-1}$ (\texttt{1d28\_0.4v})) produces an identical differential shift.

\begin{figure}
  \centering
  \includegraphics[width=\linewidth]{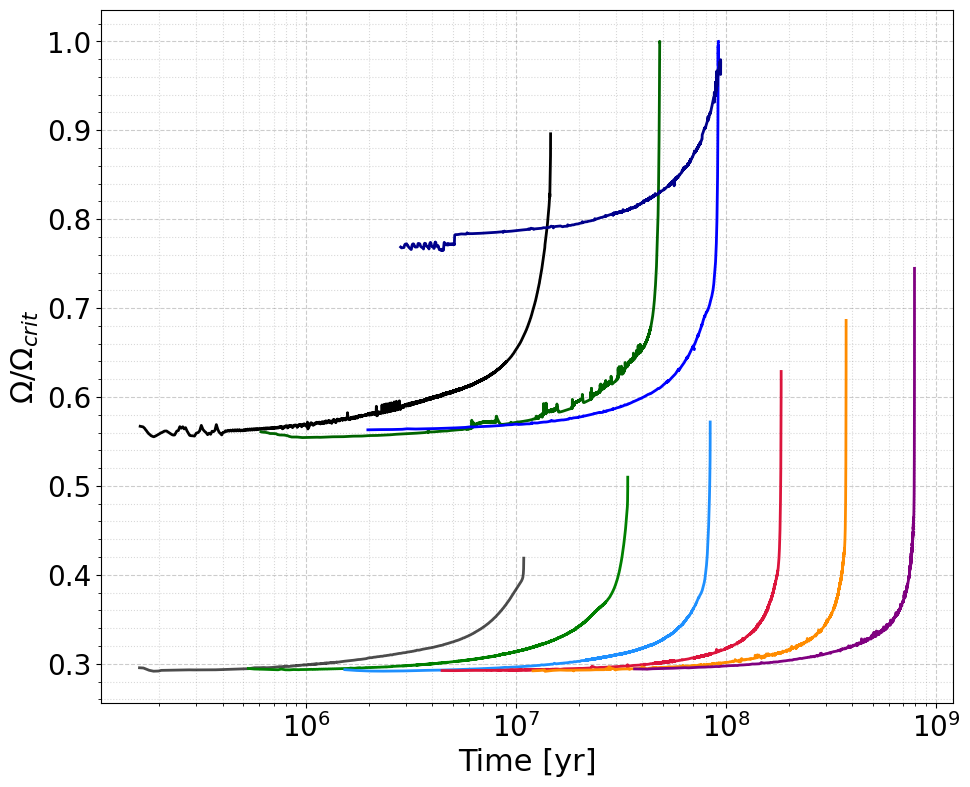}
  \caption{Surface $\Omega/\Omega_{\rm crit}$ versus time for 20 (\textit{bottom}), 40 (\textit{middle}), and 60\% (\textit{top}) \,$v_{\rm crit}$\ models at several WIMPs density with a color scheme similar to \ref{fig:HRD_rot-all}.}
  \label{fig:Om-log}
\end{figure}

Figure~\ref{fig:Om-log} demonstrates that dark‐matter longevity enhances the surface spin‐up driven by internal angular‐momentum transport.  
At $\rho_\chi=6.3\times10^{9}$ GeV cm$^{-3}$ the 40 \%model (\texttt{63d29\_0.4v}) attains $\Omega/\Omega_{\rm crit}=0.99$ at the end of MS, whereas the same model without DM has reached only 0.9 by that age. The higher DM density extends the Kelvin–Helmholtz spin‐up phase to $\sim700$ Myr, doubling the time available for meridional circulation to transfer core angular momentum outward. In the 60\% sequence the stellar surface meets the break-up limit at $\simeq90$ Myr, long before the core hydrogen fraction falls below 0.6; the computation therefore halts. This behaviour implies that any Pop III star born with $v\gtrsim0.5\,v_{\rm crit}$ in a DM halo of $\rho_\chi \sigma_\text{SD} = 1\cdot 10^{-28}$ $\text{GeV}\cdot\text{cm}^{-1}$ will likely reach critical rotation well inside its DM-supported phase.

\begin{figure}
  \centering
  \includegraphics[width=\linewidth]{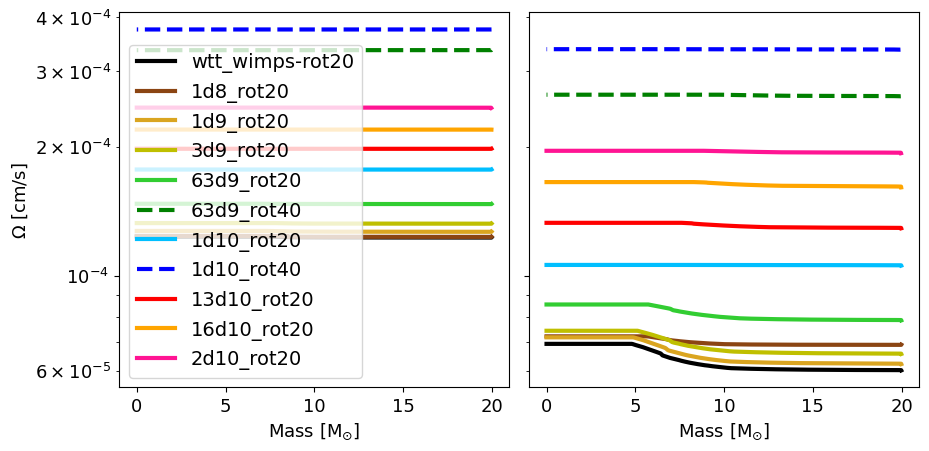}
  \caption{Angular-velocity profile $\Omega(m)$ at mid-MS \textit{left},  and MS end \textit{right} for different WIMPs densities and initial velocities.}
  \label{fig:Om_prof_all}
\end{figure}

Internal rotation profiles in Fig.~\ref{fig:Om_prof_all} show how DM-induced longevity enforces nearly solid‐body rotation. \\
 All models present a constant rotation profile at the mid-MS ($X_{c}=0.37$), while at the end-MS the contrast between center and surface is $\simeq 15$\% for $\rho_\chi\sigma_{\rm SD} < 1.0\times10^{-28}$ GeV cm$^{-1}$ and $\lesssim 1$\% for the all other models, highlighting that dark-matter energy input, by prolonging the MS by several orders of magnitude, gives rotation enough time to homogenize the interior angular momentum.\\

\subsubsection{Main-sequence lifetimes}\label{sec:rot_lifetimes}

\begin{figure}
  \centering
  \includegraphics[width=\linewidth]{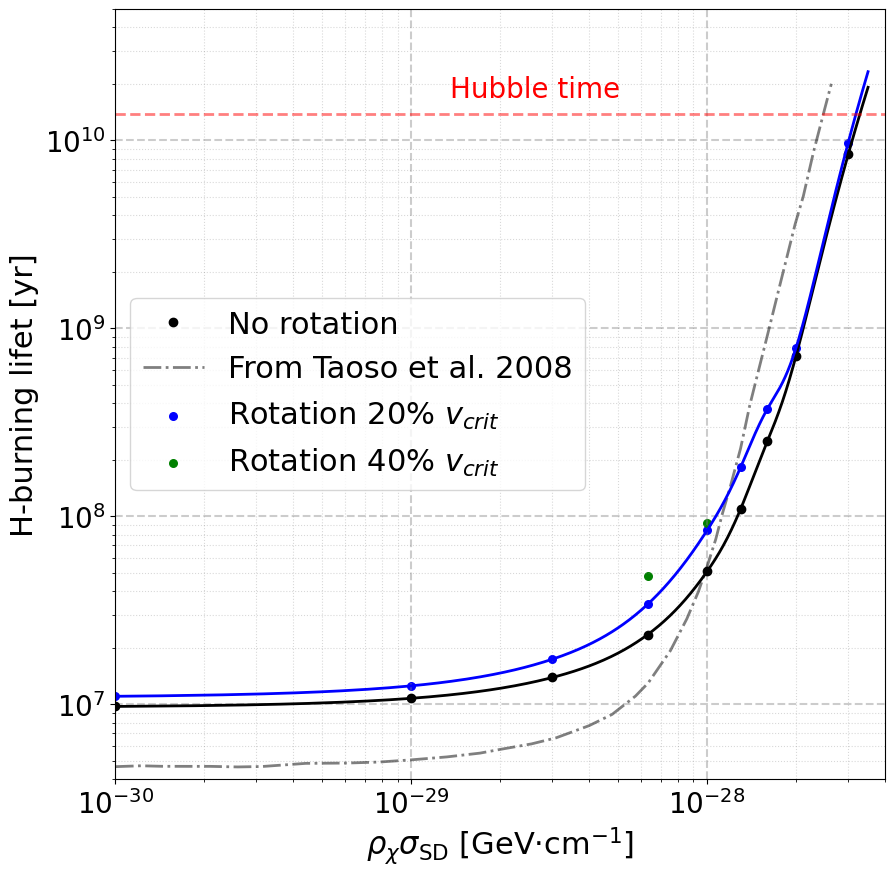}
  \caption{MS lifetime versus $\rho_\chi$ for 0 (black), 20 (blue), and 40\% $v_{\rm crit}$ (green). The grey dashed line is taken from T08, describing a static 20 M$_\odot$ star.}
  \label{fig:lifetime}
\end{figure}

Lifetimes in Fig.~\ref{fig:lifetime} underline the synergy of rotation and DM.  
At $\rho_\chi\sigma_{\rm SD}=6.3\times10^{-29}$ GeV cm$^{-1}$ the MS lasts 23.5, 34.1, and 48.4 Myr for 0, 20, and 40\% $v_{\rm crit}$, respectively; at $\rho_\chi\sigma_{\rm SD}=1.0\times10^{-28}$ GeV cm$^{-1}$ the sequence reads 50.9, 84.2, 92.0 Myr.  
The absolute gain from rotation is largest at intermediate DM supply, where nuclear burning still contributes appreciably and can benefit from fresh H mixed inward.  
For $\rho_\chi \sigma_{\rm SD} \gtrsim 2 \times 10^{-28}$ GeV cm$^{-1}$ the non-rotating lifetime already exceeds 0.7 Gyr; rotation adds only a $\sim 10$\% increment because WIMP heating so completely suppresses fusion that additional mixing has little leverage.

Repeating the 20 and 40\% sequences with $\sigma_{\rm SD}=10^{-40}$ cm$^{2}$ and proportionally larger $\rho_\chi$ reproduces the HRDs, core masses, and lifetimes to within 3\% of the values above; see Table~\ref{tab:all_data}.

Rotation reinforces the effects of dark-matter heating by (i) raising the surface luminosity at fixed $\rho_\chi$, (ii) driving the star toward near-solid-body rotation, (iii) building substantially larger helium cores, and (iv) further extending the already prolonged MS lifetime.  At sufficiently high spin and DM density the star approaches critical rotation well before core hydrogen exhaustion, a regime requiring dedicated treatment of centrifugal mass loss that lies beyond the scope of the present work.

\subsection{Chemical abundance evolution in DM–supported Pop\,III stars}\label{sec:abundances}

Figure~\ref{fig:ab-rot20} shows how dark-matter heating and rotation reshape the chemical abundances in a 20\,M$_\odot$ Pop\,III star with WIMPs and rotation. Four evolutionary checkpoints are shown: mid–H burning ($X_{c}\!\simeq\!0.37$), end of H burning ($X_{c}\!\simeq\!10^{-3}$), mid–He burning ($Y_{c}\!\simeq\!0.50$), and end of He burning ($Y_{c}\!\simeq\!10^{-3}$). The center is fully convective so the central abundance, noted $A_c$ with $A$ an arbitrary element, is the same across the core. $A_s$ is taken at the stellar surface.  

\subsubsection{H and He abundances}

For a star without dark matter and rotation the surface abundances remain constant during evolution with $X_s = 0.75$ and $Y_s = 0.25$. Table~\ref{tab:all_abundances} shows that these abundances do not vary between models without WIMPs, either with rotation (\texttt{wtt\_0.2v}) or without rotation (\texttt{wtt}), and the model with WIMPs but no rotation (\texttt{1d28\_new}). The only differences appear in the model combining WIMPs and rotation, \texttt{1d30\_0.2v}. In the middle of the MS the gradient from core to surface decreases from $\Delta X = \Delta Y = 0.38$ to $\Delta X = \Delta Y = 0.20$, caused by higher He and lower H surface abundances due to homogenization.  

At higher spin (\texttt{1d28\_0.4v}, not in Table~\ref{tab:all_abundances}) rotation alone nearly achieves full homogenization, with $\Delta X = \Delta Y = 0.10$ in the middle of MS ($X_s = 0.47$, $Y_s = 0.53$). At higher WIMP density (\texttt{3d28\_0.2v}, not in Table~\ref{tab:all_abundances}) full homogenization is almost reached, with $\Delta X = \Delta Y = 0.02$ in the middle of MS ($X_s = 0.40$, $Y_s = 0.60$). Adding dark matter to rotation extends the period over which the star remains mixed. Observable effects include helium-enriched surfaces and weaker core–envelope composition contrasts, both of which alter ionizing spectra and late-stage yields.  

\subsubsection{CNO abundance patterns and primary nitrogen}\label{sec:CNO}

\begin{figure}
    \centering
    \includegraphics[width=1.\linewidth]{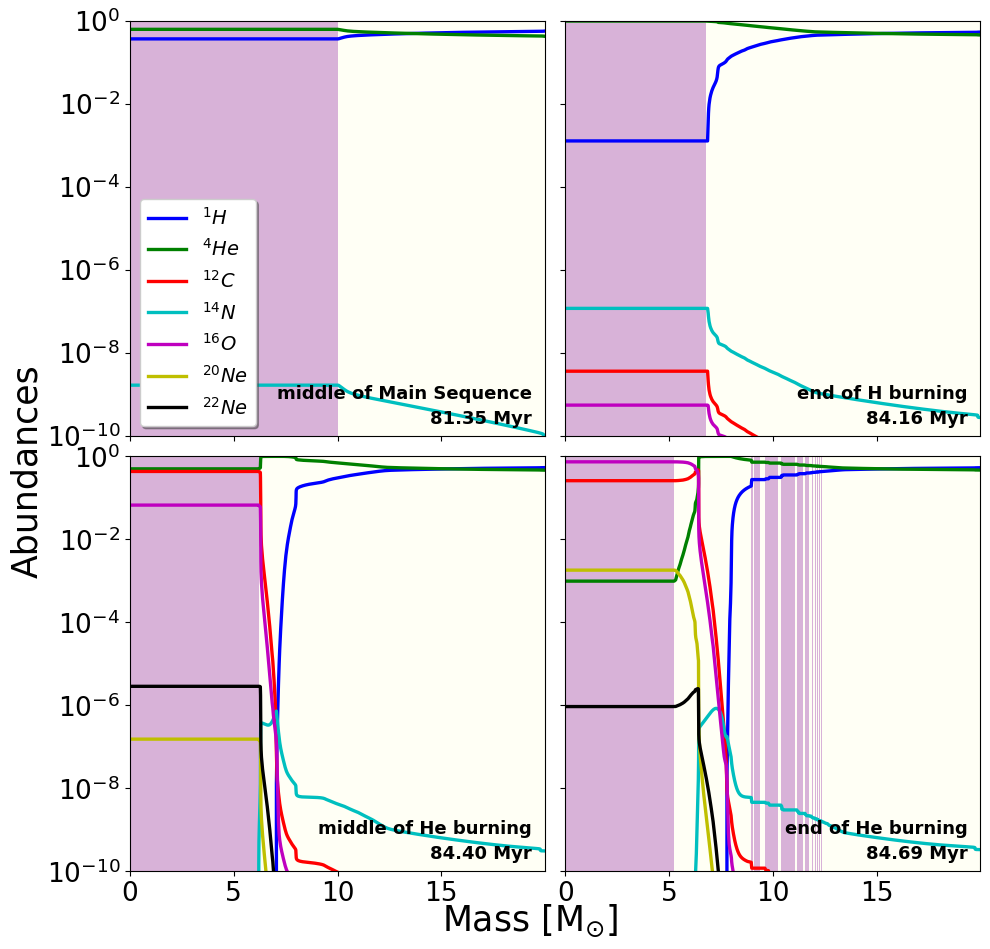}
    \caption{Abundances of the main elements inside a 20 M$_\odot$ with $\rho_\chi \sigma_\text{SD} = 1\cdot 10^{-28}$ $\text{GeV}\cdot\text{cm}^{-1}$ at 20\% of $v_{\text{crit}}$ initially.}
    \label{fig:ab-rot20}
\end{figure}

Figure~\ref{fig:ab-rot20} and Table~\ref{tab:all_abundances} show the mass fractions of the main CNO isotopes ($^{12}$C, $^{14}$N, $^{16}$O), Neon is not included. The figure shows a sharp CNO spike confined inside the helium core ($m/M \lesssim 1/3$). During the MS the abundances remain below $10^{-8}$ for $X_{^{12}\mathrm{C}}$, $10^{-7}$ for $X_{^{14}\mathrm{N}}$, and $10^{-9}$ for $X_{^{16}\mathrm{O}}$, which is consistent across all initial conditions. Models without rotation keep metal-free envelopes, while model \texttt{wtt\_0.2v} reaches $X_{^{12}\mathrm{C}} = 2.8\times10^{-14}$, $X_{^{14}\mathrm{N}} = 6.6\times10^{-12}$, and $X_{^{16}\mathrm{O}} = 1.3\times10^{-13}$ at the end of the MS. Adding DM with rotation, in model \texttt{1d28\_0.2v}, raises the surface abundances by a factor of $\approx 10^2$ at the end of the MS, giving $X_{^{12}\mathrm{C}} = 1.33\times10^{-12}$, $X_{^{14}\mathrm{N}} = 2.7\times10^{-10}$, and $X_{^{16}\mathrm{O}} = 4.8\times10^{-12}$.  

As with H and He, stronger rotation and higher WIMP density enhance mixing and increase the CNO surface abundances. In model \texttt{1d28\_0.4v} the values are $X_{^{12}\mathrm{C}} = 4.4\times10^{-11}$, $X_{^{14}\mathrm{N}} = 6.9\times10^{-9}$, and $X_{^{16}\mathrm{O}} = 9.6\times10^{-11}$. In model \texttt{3d28\_0.2v} the values are $X_{^{12}\mathrm{C}} = 2.6\times10^{-11}$, $X_{^{14}\mathrm{N}} = 4.3\times10^{-9}$, and $X_{^{16}\mathrm{O}} = 6.3\times10^{-11}$.  

Higher rotation mixes the heavier elements ($^{12}$C, $^{14}$N, $^{16}$O) more efficiently than higher WIMP density, which instead enhances the mixing of H and He by producing lower gradients for these elements.  

\subsubsection{Supernova remnant yields}

We do not push the computations until the end of core Si burning to better produce remnant masses, and we are aware that physics of calculating remnant masses is uncertain. However, we can nonetheless provide a first order estimate on the remnant masses using the two formulations widely available in the literature. First one is \cite{Maeder-1992}, where they find a remnant mass of 2.14 M$_\odot$ for an initial non rotating star of 20 M$_\odot$ (Table 4). They computed this mass by considering that all the skin layers have been ejected by the supernova explosion. Second one is \cite{Patton-2022}, they use non rotating models but take binary interactions into account. They find a mass of $\approx$ 5.7 M$_\odot$ (Figure 1). The discrepancy between the two values comes from different methodology to determine the remnant mass, and highlight the difficulty in estimating these values.

In this work, we computed the mass of each element expected in the future Supernovae Remnant (SNR) yields. For this we integrated the mass at the end of He burning ($Y_{c}\!\simeq\!10^{-3}$) from the edge of the helium core ($X_{^{14}\mathrm{C}} < 10^{-3}$) to the surface for each element. The results are summarized in Table~\ref{tab:integ_yields}. The integrated masses of H and He are $\sim 9$\,M$_\odot$ and $\sim 6$\,M$_\odot$, except for model \texttt{1d28\_0.2v} where the values are $\sim 5$\,M$_\odot$ and $\sim 8$\,M$_\odot$. This corresponds to an external layer that is 45\% poorer in H and 33\% richer in He. Nitrogen follows a similar trend, with a mass five times higher for \texttt{1d28\_0.2v}. Oxygen yields are two orders of magnitude larger in rotating stars, \texttt{wtt\_0.2v} and \texttt{1d28\_0.2v} ($>10^{-6}$ compared to $\sim 3\times10^{-8}$), and in \texttt{1d28\_0.2v} they are eight times higher than in the DM-free rotating model \texttt{wtt\_0.2v}.  

For the other elements the trends are less clear. The highest carbon yield is in model \texttt{wtt} ($7.3\times10^{-5}$) and the lowest is in \texttt{wtt\_0.2v} ($1.8\times10^{-5}$). The trend reverses in DM models, but the difference is negligible. Neon yields are one order of magnitude higher in \texttt{1d28\_0.2v} ($4.7\times10^{-10}$) compared to \texttt{wtt\_0.2v} ($2.7\times10^{-11}$), while they remain relatively similar in the static models \texttt{wtt} and \texttt{1d28} ($9.6\times10^{-10}$ and $2.7\times10^{-10}$).  

If we consider that the remnant mass will be the remaining helium core while the surface layers are ejected by the supernova explosion (similarly as in \cite{Maeder-1992}), we get 4.8 M$_\odot$ for \texttt{wtt}, 5.1 M$_\odot$ for \texttt{1d28\_new}, 5.0 M$_\odot$ for \texttt{wtt\_0.2v}, and 7 M$_\odot$ for \texttt{1d28\_0.2v}. The masses are more similar to \cite{Patton-2022} than \cite{Maeder-1992}, and with a remnant mass $\approx$ 40\% higher for the star with DM and rotation. However, these results are to be taken carefully as we do not compute the models until presupernova stage.

In conclusion, the chemical composition of SNRs may provide a way to probe DM in Pop\,III stars. The rotating model with WIMPs is the most promising as it shows increased He, N, and O yields together with lower H. Carbon and neon are not good tracers because their SNR abundances differ little from DM-free models. The remnant masses could be a way to detect those objects as well, albeit more work needs to be done regarding this topic.

\subsubsection{DM–induced mixing of H and He}\label{sec:DM_HHe}

\begin{table*}
    \centering
    \caption{Abundances at the core and at the surface as well as the gradient between both abundances ($\Delta A = |A_c - A_s|$) for different elements and different initial conditions of a 20 M$_\odot$ Pop III star.}
    {
\begin{tabular}{|c|ccc||ccc||ccc||ccc|}
    \hline
      & \multicolumn{3}{c}{\texttt{wtt}} & \multicolumn{3}{c}{\texttt{1d28\_new}} & \multicolumn{3}{c}{\texttt{wtt\_0.2v}} & \multicolumn{3}{c}{\texttt{1d10\_0.2v}} \\ 
      & $A_c$ & $A_s$ & $\Delta A$ & $A_c$ & $A_s$ & $\Delta A$ & $A_c$ & $A_s$ & $\Delta A$& $A_c$ & $A_s$ & $\Delta A$\\ 
    \hline \hline
    Middle of MS & & & & & & & & & & & & \\
    $^1$H   & 0.37 & 0.75 & 0.38   &   0.37 & 0.75 & 0.38   &   0.37 & 0.75 & 0.38   &   0.37 & 0.57 & 0.20 \\
    $^4$He  & 0.63 & 0.25 & 0.38   &   0.63 & 0.25 & 0.38   &   0.63 & 0.25 & 0.38   &   0.63 & 0.42 & 0.20 \\
    $^{12}$C   & 0.00 & 0.00 & 0.00   &   0.00 & 0.00 & 0.00   &   0.00 & 0.00 & 0.00   &   0.00 & 0.00 & 0.00 \\
    $^{14}$N   & 0.00 & 0.00 & 0.00   &   0.00 & 0.00 & 0.00   &   0.00 & 0.00 & 0.00   &   0.00 & 0.00 & 0.00 \\
    $^{16}$O   & 0.00 & 0.00 & 0.00   &   0.00 & 0.00 & 0.00   &   0.00 & 0.00 & 0.00   &   0.00 & 0.00 & 0.00 \\

    \hline \hline
    End of H burning & & & & & & & & & & & & \\
    $^1$H   & 0.00 & 0.75 & 0.75   &   0.00 & 0.75 & 0.75   &   0.00 & 0.75 & 0.75   &   0.00 & 0.54 & 0.54 \\
    $^4$He  & 1.00 & 0.25 & 0.75   &   1.00 & 0.25 & 0.75   &   1.00 & 0.25 & 0.75   &   1.00 & 0.46 & 0.54 \\
    $^{12}$C   & 0.00 & 0.00 & 0.00   &   0.00 & 0.00 & 0.00   &   0.00 & 0.00 & 0.00   &   0.00 & 0.00 & 0.00 \\
    $^{14}$N   & 0.00 & 0.00 & 0.00   &   0.00 & 0.00 & 0.00   &   0.00 & 0.00 & 0.00   &   0.00 & 0.00 & 0.00 \\
    $^{16}$O   & 0.00 & 0.00 & 0.00   &   0.00 & 0.00 & 0.00   &   0.00 & 0.00 & 0.00   &   0.00 & 0.00 & 0.00 \\

    \hline\hline
    Middle of He burning & & & & & & & & & & & & \\
    $^1$H   & 0.00 & 0.75 & 0.75   &   0.00 & 0.74 & 0.74   &   0.00 & 0.75 & 0.75   &  0.00 & 0.54 & 0.53\\
    $^4$He  & 0.50 & 0.25 & 0.25   &   0.50 & 0.26 & 0.24   &   0.50 & 0.25 & 0.25   &  0.50 & 0.47 & 0.36\\
    $^{12}$C   & 0.44 & 0.00 & 0.44   &   0.44 & 0.00 & 0.44   &   0.44 & 0.00 & 0.44   &  0.43 & 0.00 & 0.43 \\
    $^{14}$N   & 0.00 & 0.00 & 0.00   &   0.00 & 0.00 & 0.00   &   0.00 & 0.00 & 0.00   &  0.00 & 0.00 & 0.00 \\
    $^{16}$O   & 0.06 & 0.00 & 0.06   &   0.06 & 0.00 & 0.06   &   0.06 & 0.00 & 0.06   &  0.07 & 0.00 & 0.07 \\

    \hline\hline
    End of He burning & & & & & & & & & & & &\\
    $^1$H   & 0.00 & 0.75 & 0.75   &   1.00 & 0.74 & 0.74   &   0.00 & 0.75 & 0.75   &   0.00 & 0.53 & 0.53 \\
    $^4$He  & 0.00 & 0.25 & 0.25   &   0.00 & 0.26 & 0.26   &   0.00 & 0.25 & 0.25   &   0.00 & 0.47 & 0.47 \\
    $^{12}$C   & 0.30 & 0.00 & 0.30   &   0.29 & 0.00 & 0.29   &   0.29 & 0.00 & 0.29   &   0.26 & 0.00 & 0.26 \\
    $^{14}$N   & 0.00 & 0.00 & 0.00   &   0.00 & 0.00 & 0.00   &   0.00 & 0.00 & 0.00   &   0.00 & 0.00 & 0.00 \\
    $^{16}$O   & 0.70 & 0.00 & 0.70   &   0.71 & 0.00 & 0.71   &   0.71 & 0.00 & 0.71   &   0.74 & 0.00 & 0.74 \\
    \hline
    \end{tabular}
    }
    \label{tab:all_abundances}
\end{table*}

\begin{table}
    \centering
    \caption{Mass fraction for H, He, C, N, O and Ne outside the Helium core, supposedly ending up in the Supernovae remnant}
    {
\begin{tabular}{|c|cccc|}
    \hline
     [M$_\odot$] & \texttt{wtt} & \texttt{1d28\_new} & \texttt{wtt\_0.2v} & \texttt{1d28\_0.2v} \\ 
    \hline
    H   &  9.0 &  8.7 & 8.7 & 5.1 \\
    He & 6.2  &  6.2 & 6.3 & 7.9 \\
    C & $7.3\cdot10^{-5}$ &  $4.3\cdot10^{-5}$ & $1.8\cdot10^{-5}$ & $5.8\cdot10^{-5}$ \\
    N   & $1.3\cdot10^{-7}$ & $1.1\cdot10^{-7}$ & $1.3\cdot10^{-7}$ & $5.2\cdot10^{-7}$ \\
    O   & $3.8\cdot10^{-8}$ & $2.4\cdot10^{-8}$ & $1.7\cdot10^{-6}$ & $8.1\cdot10^{-6}$ \\
    Ne   & $9.6\cdot10^{-10}$ & $2.7\cdot10^{-10}$ & $4.9\cdot10^{-11}$ & $4.7\cdot10^{-10}$ \\
    \hline
\end{tabular}
    }
    \label{tab:integ_yields}
\end{table}

The integrated masses of H and He are $\sim 9$ M$_\odot$ and $\sim 6$ M$_\odot$ respectively, except for \texttt{1d28\_0.2v} where the integrated masses are $\sim 5$ M$_\odot$ and $\sim 8$ M$_\odot$ respectively, i.e. the external part of the star is 45\% poorer in H and 33\% richer in He. N has a similar trend with a mass 5 times higher for \texttt{1d28\_0.2v}. The oxygen yields are 2 orders of magnitude more massive for rotating stars, \texttt{wtt\_0.2v} and \texttt{1d28\_0.2v} ($ >10^{-6}$ compare to $\sim 3 \cdot 10^{-8}$), especially for \texttt{1d28\_0.2v} which is 8 times higher than the DM-less rotating star \texttt{wtt\_0.2v}. \\
For the two other elements, finding a general trend is more difficult. The highest carbon yields is present for \texttt{wtt} ($7.3\cdot10^{-5}$) while the lowest is present for \texttt{wtt\_0.2v} ($1.8\cdot10^{-5}$). A trend which is opposite when the star is surrounded by DM, but with a negligible difference.
The Neon yield is one order of magnitude higher for \texttt{1d28\_0.2v} model  ($4.7\cdot10^{-10}$) compare to \texttt{wtt\_0.2v} ($2.7\cdot10^{-11}$), while they are relatively similar for the static stars, \texttt{wtt} and \texttt{1d28} ($9.6\cdot10^{-10}$ and $2.7\cdot10^{-10}$).

In conclusion, a way to look for DM in PopIII stars is to look at the chemical composition of the SNR from those stars. The rotating model with WIMPs is the most promising as it exhibits an increase in He, N and O, while decreasing for H. C and Ne do not seem to be good tracers to detect stars surrounded by DM as their proportion in SNR do not differ significantly from the DM-less models. 

\section{Discussion}%
\label{sec:discussion}

The 20 M$_\odot$ mass prescription chosen in this work is a suitable candidate for type II supernova (core collapse supernova). Additionally, 20 M$_\odot$ is a classical choice when looking at Initial Mass Functions (IMFs) not only for in local universe but also for high redshift universe, this means if stars powered by dark matter exist, it would be a much higher chance to observe such a star ($\xi(m)\Delta m \approx 6\times10^{-3}$). Lastly, the lifetime of this star without any dark matter is long enough for any rotation processes to be significant ($\approx$ 9 Myr for static and $\approx$ 10 Myr for rotating star \citep{Ekstrom2012}). As an example, if we were to take a 60 M$_\odot$ star, its lifespan would decrease to a few million years and the chemical mixing would be less significant. A rotating 20 M$_\odot$ star, whose internal structure is composed of a convective core and a radiative envelope, does not produce a homogeneous evolution. However, adding dark matter to this star, we indeed find homogeneous evolution due to lifetime extension, and allowing the transport of chemicals even through the radiative zones. In addition, the likelihood of finding a 20 M$_\odot$ star, despite there being hints of a top heavy (IMF) for PopIII stars, is much higher than, for instance, 60 M$_\odot$ stars \citep{Chen2020}. This makes the 20 M$_\odot$ model a viable case for comparison in this work.

WIMP annihilation reshapes the internal composition of our $20\,M_\odot$ grid. The non-rotating control model builds a steep H–He discontinuity with $\Delta X\simeq 0.38$ at $m/M \simeq 0.22$ and retains a pristine envelope at $X\approx0.75$. When DM heating is included with $v_{\mathrm{ini}}=0.40\,v_{\mathrm{crit}}$ the star evolves quasi-homogeneously and the gradient falls to $\Delta X\simeq0.10$ at H exhaustion. The surface nitrogen mass fraction rises from $6.6\times10^{-12}$ to $2.7\times10^{-10}$ at the end of MS. These values exceed the primary nitrogen reported by \citet{Yoon2008} for the same mass. The non-rotating DM case shows weaker enrichment and thus confirms that DM and rotation together drive the extreme CNO yields.\\
Lifetimes scale with the product $\rho_\chi\sigma_{\mathrm{SD}}$ as anticipated by \citet{taoso_dark_2008}. The reference model without DM leaves the main sequence after $9.6\,$Myr. At $\rho_\chi=6.3\times10^{9}\,$GeV\,cm$^{-3}$ the lifetime doubles to $23.5\,$Myr, matching the $20\,M_\odot$ result of \citet{Scott2009} to within $5\%$. Raising the density to $1.0\times10^{10}$ extends the lifetime to $50.9\,$Myr, a factor $5.3$ that agrees with the $4.9$ reported by \citet{Casanellas2011}. Beyond $2\times10^{10}$ our models fail to exhaust hydrogen within a Hubble time, reproducing the threshold identified by \citet{taoso_dark_2008}. Switching to $\sigma_{\mathrm{SD}}=10^{-40}\,$cm$^{2}$ and scaling $\rho_\chi$ by $100$ recovers identical lifetimes, confirming that $\tau_{\mathrm{MS}}$ depends only on the capture product.

Rotation changes the outcome only when DM prolongs the nuclear timescale. A $v_{\mathrm{ini}}=0.40\,v_{\mathrm{crit}}$ model without DM ends core-H burning with $\Omega_{\mathrm{core}}/\Omega_{\mathrm{surf}}\simeq3$ and $M_{\mathrm{He}}=6.8\,M_\odot$  .  
Introducing $\rho_\chi=1.0\times10^{10}\,\mathrm{GeV\,cm^{-3}}$ lowers the ratio to $1.08$ (8 \%) and raises the core mass to $13.3\,M_\odot$  .  
These values fulfil the quasi-chemically homogeneous condition $\tau_{\mathrm{mix}}<\tau_{\mathrm{nuc}}$ proposed by \citet{Yoon2008}.  
With DM, the same state appears at $0.20\,v_{\mathrm{crit}}$.  This halves the CHE threshold compared with the $\gtrsim0.50\,v_{\mathrm{crit}}$ requirement in DM-free Pop III models.  
We conclude that WIMP energy support, rather than centrifugal mixing alone, dictates the onset of chemical homogeneity in primordial massive stars.

Figure~\ref{fig:core_vs_rhochi} shows the central temperature $T_c$ versus $\rho_\chi\sigma_{\rm SD}$ for a rotating 20\,M$_\odot$ Pop\,III model with WIMP annihilation, with stages defined by $X_c$ and dashed reference curves from T08. At the ZAMS the WIMP model is $26\%$ cooler and $55\%$ larger with $T_{\rm eff}\!\approx\!50$\,kK and $R\!\approx\!2.8\,R_\odot$ compared to the DM–free model with $T_{\rm eff}\!\approx\!63$\,kK and $R\!\approx\!1.8\,R_\odot$. This inflation follows from annihilation heating which supplies part of the luminosity and lowers the nuclear rate and $T_c$, which in turn raises the core entropy. At later times the annihilation rate drops and static models differ by only $2$–$8\%$ in radius and by less than $3\%$ in temperature so their structures converge. In the rotating WIMP model the trend reverses after the early MS and the radius becomes smaller by up to $17\%$ while $T_{\rm eff}$ becomes higher by $14$–$20\%$. Rotation maintains strong mixing and flattens the $\mu$ gradient which enriches the envelope in He and lowers the opacity. Therefore, the $T_{\rm eff}$ rises at fixed luminosity and the star contracts. The core remains cooler throughout, because mixing spreads the annihilation power and throttles nuclear burning.

\section{Conclusions}\label{sec:conclusions}

We have computed the first grid of $20\,M_\odot$ Population III models that self-consistently include WIMP capture, annihilation, and rotation.  The grid spans six ambient DM densities ($10^{8}$–$3\times10^{10}\,\mathrm{GeV\,cm^{-3}}$) and three initial spins ($0$, $0.20$, $0.40\,v_{\mathrm{crit}}$), allowing a controlled comparison of non-rotating, rotating, DM-free, and DM-rich tracks.

\begin{itemize}
  \item \textit{Main-sequence longevity.}  
        WIMP annihilation lengthens the lifetime of a $20\,M_\odot$ Pop III star from $\sim10$ Myr to gigayear scales once the capture product exceeds $\rho_\chi\sigma_{\mathrm{SD}}\!\sim\!2\times10^{-28}$ GeV cm$^{-1}$.  
        The vastly slower nuclear clock lets rotation and mixing act for orders of magnitude longer, reshaping subsequent feedback on the host halo.\\

  \item \textit{Rotation driven toward solid body and helium-core growth.}  
        In DM-rich tracks a model born at $0.40\,v_{\mathrm{crit}}$ ends core-H burning with only a few-percent shear and a helium core roughly twice the mass of its DM-free counterpart.  
        Near-rigid rotation suppresses shear instabilities yet sustains meridional circulation, enhancing fuel transport and setting the stage for energetic late phases.\\

  \item \textit{Central structure stabilisation.}  
        Continuous DM heating holds $T_{\mathrm c}$ and $\rho_{\mathrm c}$ nearly constant during most of hydrogen burning, delaying Kelvin–Helmholtz contraction until spin-dependent capture ceases.  
        This stabilisation maintains a low-density core that favours efficient mixing and prolongs the homogeneous phase.\\

  \item \textit{Chemically homogeneous evolution at moderate spin.}  
        DM support lengthens $\tau_{\mathrm{nuc}}$ so that Eddington–Sweet circulation overwhelms it even at $0.20\,v_{\mathrm{crit}}$, while DM-free stars need $\gtrsim0.50\,v_{\mathrm{crit}}$.  
        Quasi-homogeneity flattens the H–He profile, keeps fresh fuel in the core, and produces progenitors suitable for collapsar-type explosions.\\

  \item \textit{Surface H/He and CNO pattern with primary nitrogen.} 
        Rotation and DM mixing lowers the surface hydrogen to $X\!\sim\!0.53$, raises helium to $Y\!\sim\!0.47$, and boosts $^{12}$C, $^{14}$N and $^{16}$O by two orders of magnitude relative to classical Pop III stars\\

  \item \textit{Spectral signatures.}
        The best candidate to detect DM in PopIII stars are rotating ones surrounded by DM. Indeed, they live longer than their static counterparts and will produce SNR rich in He, N and O and poor in H, related to modified outer layer abundances due to more efficient mixing. Additionally, these stars begin their evolution as cool and bloated objects at the ZAMS, but become hotter and more compact throughout the evolution when compared to classical baryonic PopIII stars.\\

\end{itemize}

Future work will expand the mass range to $40$–$120\,M_\odot$ and probe halo densities above $3\times10^{10}\,\mathrm{GeV\,cm^{-3}}$.  Spin-independent capture on metals will be added to follow late evolutionary phases.  Magnetic torques and mass loss will be coupled to test whether DM-supported fast rotators reach critical spin and launch chemically enriched winds.  Synthetic spectra across the UV–IR range will quantify N III and He II diagnostics and calibrate ionising photon budgets for reionisation models.  Yield calculations will track $^{22}$Ne and heavy-element production and compare with N-rich extremely metal-poor stars.  Finally, models of $10^{3}$–$10^{5}\,M_\odot$ protostars with sustained DM heating will explore the formation of supermassive dark stars and the seeds of early quasars, which could be observed by the JWST (see \cite{Ilie-2023b} for an analysis of recent observations of quasars that could be Dark Stars product).

\begin{acknowledgements}
The authors would like to thank Dr. Marco Taoso and Prof. Georges Meynet for their valuable comments and feedback on the aspects of DM annihilation and rotation. Anaïs Pauchet would like to thank her supervisor Pr. Stefanie Walch for her supervision and encouragement towards the project. AP was supported by the Deutsche Forschungsgemeinschaft (DFG, German Research
Foundation) – Project-ID 500700252 – SFB 1601. DN was supported by the Swiss National Science Fund (SNSF) Postdoctoral Fellowship, grant number: P500-2235464 and by the Virginia Initiative of Theoretical Astronomy (VITA) Fellowship. 
\end{acknowledgements}

\bibliographystyle{aa}
\bibliography{biblio}

@ARTICLE{Aprile_2017,
       author = {{Aprile}, E. and {Aalbers}, J. and {Agostini}, F. and {Alfonsi}, M. and {Amaro}, F.~D. and {Anthony}, M. and {Arneodo}, F. and {Barrow}, P. and {Baudis}, L. and {Bauermeister}, B. and {Benabderrahmane}, M.~L. and {Berger}, T. and {Breur}, P.~A. and {Brown}, A. and {Brown}, A. and {Brown}, E. and {Bruenner}, S. and {Bruno}, G. and {Budnik}, R. and {B{\"u}tikofer}, L. and {Calv{\'e}n}, J. and {Cardoso}, J.~M.~R. and {Cervantes}, M. and {Cichon}, D. and {Coderre}, D. and {Colijn}, A.~P. and {Conrad}, J. and {Cussonneau}, J.~P. and {Decowski}, M.~P. and {de Perio}, P. and {di Gangi}, P. and {di Giovanni}, A. and {Diglio}, S. and {Eurin}, G. and {Fei}, J. and {Ferella}, A.~D. and {Fieguth}, A. and {Fulgione}, W. and {Gallo Rosso}, A. and {Galloway}, M. and {Gao}, F. and {Garbini}, M. and {Gardner}, R. and {Geis}, C. and {Goetzke}, L.~W. and {Grandi}, L. and {Greene}, Z. and {Grignon}, C. and {Hasterok}, C. and {Hogenbirk}, E. and {Howlett}, J. and {Itay}, R. and {Kaminsky}, B. and {Kazama}, S. and {Kessler}, G. and {Kish}, A. and {Landsman}, H. and {Lang}, R.~F. and {Lellouch}, D. and {Levinson}, L. and {Lin}, Q. and {Lindemann}, S. and {Lindner}, M. and {Lombardi}, F. and {Lopes}, J.~A.~M. and {Manfredini}, A. and {Mari{\c{s}}}, I. and {Marrod{\'a}n Undagoitia}, T. and {Masbou}, J. and {Massoli}, F.~V. and {Masson}, D. and {Mayani}, D. and {Messina}, M. and {Micheneau}, K. and {Molinario}, A. and {Mor{\^a}}, K. and {Murra}, M. and {Naganoma}, J. and {Ni}, K. and {Oberlack}, U. and {Pakarha}, P. and {Pelssers}, B. and {Persiani}, R. and {Piastra}, F. and {Pienaar}, J. and {Pizzella}, V. and {Piro}, M. -C. and {Plante}, G. and {Priel}, N. and {Rauch}, L. and {Reichard}, S. and {Reuter}, C. and {Riedel}, B. and {Rizzo}, A. and {Rosendahl}, S. and {Rupp}, N. and {Saldanha}, R. and {Dos Santos}, J.~M.~F. and {Sartorelli}, G. and {Scheibelhut}, M. and {Schindler}, S. and {Schreiner}, J. and {Schumann}, M. and {Scotto Lavina}, L. and {Selvi}, M. and {Shagin}, P. and {Shockley}, E. and {Silva}, M. and {Simgen}, H. and {Sivers}, M.~V. and {Stein}, A. and {Thapa}, S. and {Thers}, D. and {Tiseni}, A. and {Trinchero}, G. and {Tunnell}, C. and {Vargas}, M. and {Upole}, N. and {Wang}, H. and {Wang}, Z. and {Wei}, Y. and {Weinheimer}, C. and {Wulf}, J. and {Ye}, J. and {Zhang}, Y. and {Zhu}, T. and {Xenon Collaboration}},
        title = "{First Dark Matter Search Results from the XENON1T Experiment}",
      journal = {\prl},
         year = 2017,
        month = nov,
       volume = {119},
       number = {18},
          eid = {181301},
        pages = {181301},
          doi = {10.1103/PhysRevLett.119.181301},
archivePrefix = {arXiv},
       eprint = {1705.06655},
 primaryClass = {astro-ph.CO},
       adsurl = {https://ui.adsabs.harvard.edu/abs/2017PhRvL.119r1301A}
}

@article{gould_1987b,
	title = {WEAKLY INTERACTING MASSIVE PARTICLE DISTRIBUTION IN AND EVAPORATION FROM THE SUN},
	doi = {10.1086/165652},
	language = {en},
	urldate = {2023-03-22},
	journal = {The Astrophysical Journal},
	author = {Gould, Andrew},
	month = oct,
	year = {1987},
}

@article{taoso_dark_2008,
	title = {Dark matter annihilations in {Population} {III} stars},
	issn = {1550-7998, 1550-2368},
	url = {https://link.aps.org/doi/10.1103/PhysRevD.78.123510},
	doi = {10.1103/PhysRevD.78.123510},
	language = {en},
	journal = {Physical Review D},
	author = {Taoso, Marco and Bertone, Gianfranco and Meynet, Georges and Ekström, Silvia},
	month = dec,
	year = {2008}
}

@ARTICLE{Ekstrom2012,
       author = {{Ekstr{\"o}m}, S. and {Georgy}, C. and {Eggenberger}, P. and {Meynet}, G. and {Mowlavi}, N. and {Wyttenbach}, A. and {Granada}, A. and {Decressin}, T. and {Hirschi}, R. and {Frischknecht}, U. and {Charbonnel}, C. and {Maeder}, A.},
        title = "{Grids of stellar models with rotation. I. Models from 0.8 to 120 M$_{{\ensuremath{\odot}}}$ at solar metallicity (Z = 0.014)}",
      journal = {\aap},
         year = 2012,
        month = jan,
       volume = {537},
          eid = {A146},
        pages = {A146},
          doi = {10.1051/0004-6361/201117751},
archivePrefix = {arXiv},
       eprint = {1110.5049},
 primaryClass = {astro-ph.SR},
       adsurl = {https://ui.adsabs.harvard.edu/abs/2012A&A...537A.146E}
}

@ARTICLE{Nandal2024b,
       author = {{Nandal}, Devesh and {Meynet}, Georges and {Ekstr{\"o}m}, Sylvia and {Moyano}, Facundo D. and {Eggenberger}, Patrick and {Choplin}, Arthur and {Georgy}, Cyril and {Farrell}, Eoin and {Maeder}, Andr{\'e}},
        title = "{Impact of different approaches to computing rotating stellar models. I. The case of solar metallicity}",
      journal = {\aap},
         year = 2024,
        month = apr,
       volume = {684},
          eid = {A169},
        pages = {A169},
          doi = {10.1051/0004-6361/202346979},
archivePrefix = {arXiv},
       eprint = {2312.13340},
 primaryClass = {astro-ph.SR},
       adsurl = {https://ui.adsabs.harvard.edu/abs/2024A&A...684A.169N}
}

@ARTICLE{Jungman1996,
       author = {{Jungman}, G. and {Kamionkowski}, M. and {Griest}, K.},
        title = "{Supersymmetric dark matter}",
      journal = {\physrep},
         year = 1996,
        month = mar,
       volume = {267},
        pages = {195-373},
          doi = {10.1016/0370-1573(95)00058-5},
archivePrefix = {arXiv},
       eprint = {hep-ph/9506380},
 primaryClass = {hep-ph},
       adsurl = {https://ui.adsabs.harvard.edu/abs/1996PhR...267..195J}
}

@ARTICLE{Yoon2008,
       author = {{Yoon}, Sung-Chul and {Iocco}, Fabio and {Akiyama}, Shizuka},
        title = "{Evolution of the First Stars with Dark Matter Burning}",
      journal = {\apjl},
         year = 2008,
        month = nov,
       volume = {688},
       number = {1},
        pages = {L1},
          doi = {10.1086/593976},
archivePrefix = {arXiv},
       eprint = {0806.2662},
 primaryClass = {astro-ph},
       adsurl = {https://ui.adsabs.harvard.edu/abs/2008ApJ...688L...1Y}
}

@ARTICLE{Ilie-2023,
       author = {{Ilie}, Cosmin and {Paulin}, Jillian and {Freese}, Katherine},
        title = "{Supermassive Dark Star candidates seen by JWST}",
      journal = {Proceedings of the National Academy of Science},
         year = 2023,
        month = jul,
       volume = {120},
       number = {30},
          eid = {e2305762120},
        pages = {e2305762120},
          doi = {10.1073/pnas.2305762120},
archivePrefix = {arXiv},
       eprint = {2304.01173},
 primaryClass = {astro-ph.CO},
       adsurl = {https://ui.adsabs.harvard.edu/abs/2023PNAS..12005762I}
}

@ARTICLE{freese-2016,
       author = {{Freese}, Katherine and {Rindler-Daller}, Tanja and {Spolyar}, Douglas and {Valluri}, Monica},
        title = "{Dark stars: a review}",
      journal = {Reports on Progress in Physics},
         year = 2016,
        month = jun,
       volume = {79},
       number = {6},
          eid = {066902},
        pages = {066902},
          doi = {10.1088/0034-4885/79/6/066902},
archivePrefix = {arXiv},
       eprint = {1501.02394},
 primaryClass = {astro-ph.CO},
       adsurl = {https://ui.adsabs.harvard.edu/abs/2016RPPh...79f6902F}
}

@ARTICLE{Stacy-2012,
       author = {{Stacy}, Athena and {Greif}, Thomas H. and {Bromm}, Volker},
        title = "{The first stars: mass growth under protostellar feedback}",
      journal = {\mnras},
         year = 2012,
        month = may,
       volume = {422},
       number = {1},
        pages = {290-309},
          doi = {10.1111/j.1365-2966.2012.20605.x},
archivePrefix = {arXiv},
       eprint = {1109.3147},
 primaryClass = {astro-ph.CO},
       adsurl = {https://ui.adsabs.harvard.edu/abs/2012MNRAS.422..290S}
}

@ARTICLE{Ilie-2021,
       author = {{Ilie}, Cosmin and {Levy}, Caleb and {Pilawa}, Jacob and {Zhang}, Saiyang},
        title = "{Constraining dark matter properties with the first generation of stars}",
      journal = {\prd},
         year = 2021,
        month = dec,
       volume = {104},
       number = {12},
          eid = {123031},
        pages = {123031},
          doi = {10.1103/PhysRevD.104.123031},
archivePrefix = {arXiv},
       eprint = {2009.11474},
 primaryClass = {astro-ph.CO},
       adsurl = {https://ui.adsabs.harvard.edu/abs/2021PhRvD.104l3031I}
}

@ARTICLE{Scott-2009,
       author = {{Scott}, Pat and {Fairbairn}, Malcolm and {Edsj{\"o}}, Joakim},
        title = "{Dark stars at the Galactic Centre - the main sequence}",
      journal = {\mnras},
         year = 2009,
        month = mar,
       volume = {394},
       number = {1},
        pages = {82-104},
          doi = {10.1111/j.1365-2966.2008.14282.x},
archivePrefix = {arXiv},
       eprint = {0809.1871},
 primaryClass = {astro-ph},
       adsurl = {https://ui.adsabs.harvard.edu/abs/2009MNRAS.394...82S}
}

@ARTICLE{PlanckColl-2018,
       author = {{Planck Collaboration} and {Aghanim}, N. and {Akrami}, Y. and {Ashdown}, M. and {Aumont}, J. and {Baccigalupi}, C. and {Ballardini}, M. and {Banday}, A.~J. and {Barreiro}, R.~B. and {Bartolo}, N. and {Basak}, S. and {Battye}, R. and {Benabed}, K. and {Bernard}, J. -P. and {Bersanelli}, M. and {Bielewicz}, P. and {Bock}, J.~J. and {Bond}, J.~R. and {Borrill}, J. and {Bouchet}, F.~R. and {Boulanger}, F. and {Bucher}, M. and {Burigana}, C. and {Butler}, R.~C. and {Calabrese}, E. and {Cardoso}, J. -F. and {Carron}, J. and {Challinor}, A. and {Chiang}, H.~C. and {Chluba}, J. and {Colombo}, L.~P.~L. and {Combet}, C. and {Contreras}, D. and {Crill}, B.~P. and {Cuttaia}, F. and {de Bernardis}, P. and {de Zotti}, G. and {Delabrouille}, J. and {Delouis}, J. -M. and {Di Valentino}, E. and {Diego}, J.~M. and {Dor{\'e}}, O. and {Douspis}, M. and {Ducout}, A. and {Dupac}, X. and {Dusini}, S. and {Efstathiou}, G. and {Elsner}, F. and {En{\ss}lin}, T.~A. and {Eriksen}, H.~K. and {Fantaye}, Y. and {Farhang}, M. and {Fergusson}, J. and {Fernandez-Cobos}, R. and {Finelli}, F. and {Forastieri}, F. and {Frailis}, M. and {Fraisse}, A.~A. and {Franceschi}, E. and {Frolov}, A. and {Galeotta}, S. and {Galli}, S. and {Ganga}, K. and {G{\'e}nova-Santos}, R.~T. and {Gerbino}, M. and {Ghosh}, T. and {Gonz{\'a}lez-Nuevo}, J. and {G{\'o}rski}, K.~M. and {Gratton}, S. and {Gruppuso}, A. and {Gudmundsson}, J.~E. and {Hamann}, J. and {Handley}, W. and {Hansen}, F.~K. and {Herranz}, D. and {Hildebrandt}, S.~R. and {Hivon}, E. and {Huang}, Z. and {Jaffe}, A.~H. and {Jones}, W.~C. and {Karakci}, A. and {Keih{\"a}nen}, E. and {Keskitalo}, R. and {Kiiveri}, K. and {Kim}, J. and {Kisner}, T.~S. and {Knox}, L. and {Krachmalnicoff}, N. and {Kunz}, M. and {Kurki-Suonio}, H. and {Lagache}, G. and {Lamarre}, J. -M. and {Lasenby}, A. and {Lattanzi}, M. and {Lawrence}, C.~R. and {Le Jeune}, M. and {Lemos}, P. and {Lesgourgues}, J. and {Levrier}, F. and {Lewis}, A. and {Liguori}, M. and {Lilje}, P.~B. and {Lilley}, M. and {Lindholm}, V. and {L{\'o}pez-Caniego}, M. and {Lubin}, P.~M. and {Ma}, Y. -Z. and {Mac{\'\i}as-P{\'e}rez}, J.~F. and {Maggio}, G. and {Maino}, D. and {Mandolesi}, N. and {Mangilli}, A. and {Marcos-Caballero}, A. and {Maris}, M. and {Martin}, P.~G. and {Martinelli}, M. and {Mart{\'\i}nez-Gonz{\'a}lez}, E. and {Matarrese}, S. and {Mauri}, N. and {McEwen}, J.~D. and {Meinhold}, P.~R. and {Melchiorri}, A. and {Mennella}, A. and {Migliaccio}, M. and {Millea}, M. and {Mitra}, S. and {Miville-Desch{\^e}nes}, M. -A. and {Molinari}, D. and {Montier}, L. and {Morgante}, G. and {Moss}, A. and {Natoli}, P. and {N{\o}rgaard-Nielsen}, H.~U. and {Pagano}, L. and {Paoletti}, D. and {Partridge}, B. and {Patanchon}, G. and {Peiris}, H.~V. and {Perrotta}, F. and {Pettorino}, V. and {Piacentini}, F. and {Polastri}, L. and {Polenta}, G. and {Puget}, J. -L. and {Rachen}, J.~P. and {Reinecke}, M. and {Remazeilles}, M. and {Renzi}, A. and {Rocha}, G. and {Rosset}, C. and {Roudier}, G. and {Rubi{\~n}o-Mart{\'\i}n}, J.~A. and {Ruiz-Granados}, B. and {Salvati}, L. and {Sandri}, M. and {Savelainen}, M. and {Scott}, D. and {Shellard}, E.~P.~S. and {Sirignano}, C. and {Sirri}, G. and {Spencer}, L.~D. and {Sunyaev}, R. and {Suur-Uski}, A. -S. and {Tauber}, J.~A. and {Tavagnacco}, D. and {Tenti}, M. and {Toffolatti}, L. and {Tomasi}, M. and {Trombetti}, T. and {Valenziano}, L. and {Valiviita}, J. and {Van Tent}, B. and {Vibert}, L. and {Vielva}, P. and {Villa}, F. and {Vittorio}, N. and {Wandelt}, B.~D. and {Wehus}, I.~K. and {White}, M. and {White}, S.~D.~M. and {Zacchei}, A. and {Zonca}, A.},
        title = "{Planck 2018 results. VI. Cosmological parameters}",
      journal = {\aap},
         year = 2020,
        month = sep,
       volume = {641},
          eid = {A6},
        pages = {A6},
          doi = {10.1051/0004-6361/201833910},
archivePrefix = {arXiv},
       eprint = {1807.06209},
 primaryClass = {astro-ph.CO},
       adsurl = {https://ui.adsabs.harvard.edu/abs/2020A&A...641A...6P}
}

@ARTICLE{Bertone-2005,
       author = {{Bertone}, Gianfranco and {Hooper}, Dan and {Silk}, Joseph},
        title = "{Particle dark matter: evidence, candidates and constraints}",
      journal = {\physrep},
         year = 2005,
        month = jan,
       volume = {405},
       number = {5-6},
        pages = {279-390},
          doi = {10.1016/j.physrep.2004.08.031},
archivePrefix = {arXiv},
       eprint = {hep-ph/0404175},
 primaryClass = {hep-ph},
       adsurl = {https://ui.adsabs.harvard.edu/abs/2005PhR...405..279B}
}

@ARTICLE{Bergstrom-1998,
       author = {{Bergstr{\"o}m}, Lars and {Ullio}, Piero and {Buckley}, James H.},
        title = "{Observability of {\ensuremath{\gamma}} rays from dark matter neutralino annihilations in the Milky Way halo}",
      journal = {Astroparticle Physics},
         year = 1998,
        month = aug,
       volume = {9},
       number = {2},
        pages = {137-162},
          doi = {10.1016/S0927-6505(98)00015-2},
archivePrefix = {arXiv},
       eprint = {astro-ph/9712318},
 primaryClass = {astro-ph},
       adsurl = {https://ui.adsabs.harvard.edu/abs/1998APh.....9..137B}
}

@ARTICLE{Bromm-2004,
       author = {{Bromm}, Volker and {Larson}, Richard B.},
        title = "{The First Stars}",
      journal = {\araa},
         year = 2004,
        month = sep,
       volume = {42},
       number = {1},
        pages = {79-118},
          doi = {10.1146/annurev.astro.42.053102.134034},
archivePrefix = {arXiv},
       eprint = {astro-ph/0311019},
 primaryClass = {astro-ph},
       adsurl = {https://ui.adsabs.harvard.edu/abs/2004ARA&A..42...79B}
}

@ARTICLE{Gondolo-2022,
       author = {{Gondolo}, Paolo and {Sandick}, Pearl and {Shams Es Haghi}, Barmak and {Visbal}, Eli},
        title = "{Reionization in the Light of Dark Stars}",
      journal = {\apj},
         year = 2022,
        month = aug,
       volume = {935},
       number = {1},
          eid = {11},
        pages = {11},
          doi = {10.3847/1538-4357/ac7fea},
archivePrefix = {arXiv},
       eprint = {2112.04525},
 primaryClass = {astro-ph.CO},
       adsurl = {https://ui.adsabs.harvard.edu/abs/2022ApJ...935...11G}
}

@ARTICLE{Spolyar2008,
       author = {{Spolyar}, Douglas and {Freese}, Katherine and {Gondolo}, Paolo},
        title = "{Dark Matter and the First Stars: A New Phase of Stellar Evolution}",
      journal = {\prl},
         year = 2008,
        month = feb,
       volume = {100},
       number = {5},
          eid = {051101},
        pages = {051101},
          doi = {10.1103/PhysRevLett.100.051101},
archivePrefix = {arXiv},
       eprint = {0705.0521},
 primaryClass = {astro-ph},
       adsurl = {https://ui.adsabs.harvard.edu/abs/2008PhRvL.100e1101S}
}

@ARTICLE{Freese2010,
       author = {{Freese}, Katherine and {Ilie}, Cosmin and {Spolyar}, Douglas and {Valluri}, Monica and {Bodenheimer}, Peter},
        title = "{Supermassive Dark Stars: Detectable in JWST}",
      journal = {\apj},
         year = 2010,
        month = jun,
       volume = {716},
       number = {2},
        pages = {1397-1407},
          doi = {10.1088/0004-637X/716/2/1397},
archivePrefix = {arXiv},
       eprint = {1002.2233},
 primaryClass = {astro-ph.CO},
       adsurl = {https://ui.adsabs.harvard.edu/abs/2010ApJ...716.1397F}
}

@ARTICLE{Freese2008b,
       author = {{Freese}, Katherine and {Spolyar}, Douglas and {Aguirre}, Anthony},
        title = "{Dark matter capture in the first stars: a power source and limit on stellar mass}",
      journal = {\jcap},
         year = 2008,
        month = nov,
       volume = {2008},
       number = {11},
          eid = {014},
        pages = {014},
          doi = {10.1088/1475-7516/2008/11/014},
archivePrefix = {arXiv},
       eprint = {0802.1724},
 primaryClass = {astro-ph},
       adsurl = {https://ui.adsabs.harvard.edu/abs/2008JCAP...11..014F}
}

@ARTICLE{Rindler-Daller2015,
       author = {{Rindler-Daller}, T. and {Montgomery}, M.~H. and {Freese}, K. and {Winget}, D.~E. and {Paxton}, B.},
        title = "{Dark Stars: Improved Models and First Pulsation Results}",
      journal = {\apj},
         year = 2015,
        month = feb,
       volume = {799},
       number = {2},
          eid = {210},
        pages = {210},
          doi = {10.1088/0004-637X/799/2/210},
archivePrefix = {arXiv},
       eprint = {1408.2082},
 primaryClass = {astro-ph.CO},
       adsurl = {https://ui.adsabs.harvard.edu/abs/2015ApJ...799..210R}
}

@ARTICLE{Ilie2012,
       author = {{Ilie}, Cosmin and {Freese}, Katherine and {Valluri}, Monica and {Iliev}, Ilian T. and {Shapiro}, Paul R.},
        title = "{Observing supermassive dark stars with James Webb Space Telescope}",
      journal = {\mnras},
         year = 2012,
        month = may,
       volume = {422},
       number = {3},
        pages = {2164-2186},
          doi = {10.1111/j.1365-2966.2012.20760.x},
archivePrefix = {arXiv},
       eprint = {1110.6202},
 primaryClass = {astro-ph.CO},
       adsurl = {https://ui.adsabs.harvard.edu/abs/2012MNRAS.422.2164I}
}

@ARTICLE{Zackrisson2010,
       author = {{Zackrisson}, Erik and {Scott}, Pat and {Rydberg}, Claes-Erik and {Iocco}, Fabio and {Edvardsson}, Bengt and {{\"O}stlin}, G{\"o}ran and {Sivertsson}, Sofia and {Zitrin}, Adi and {Broadhurst}, Tom and {Gondolo}, Paolo},
        title = "{Finding High-redshift Dark Stars with the James Webb Space Telescope}",
      journal = {\apj},
         year = 2010,
        month = jul,
       volume = {717},
       number = {1},
        pages = {257-267},
          doi = {10.1088/0004-637X/717/1/257},
archivePrefix = {arXiv},
       eprint = {1002.3368},
 primaryClass = {astro-ph.CO},
       adsurl = {https://ui.adsabs.harvard.edu/abs/2010ApJ...717..257Z}
}

@ARTICLE{Press1985,
       author = {{Press}, W.~H. and {Spergel}, D.~N.},
        title = "{Capture by the sun of a galactic population of weakly interacting, massive particles}",
      journal = {\apj},
         year = 1985,
        month = sep,
       volume = {296},
        pages = {679-684},
          doi = {10.1086/163485},
       adsurl = {https://ui.adsabs.harvard.edu/abs/1985ApJ...296..679P}
}

@ARTICLE{Faulkner1985,
       author = {{Faulkner}, J. and {Gilliland}, R.~L.},
        title = "{Weakly interacting, massive particles and the solar neutrino flux}",
      journal = {\apj},
         year = 1985,
        month = dec,
       volume = {299},
        pages = {994-1000},
          doi = {10.1086/163766},
       adsurl = {https://ui.adsabs.harvard.edu/abs/1985ApJ...299..994F}
}

@ARTICLE{Casanellas2011b,
       author = {{Casanellas}, Jordi and {Lopes}, Il{\'\i}dio},
        title = "{Signatures of Dark Matter Burning in Nuclear Star Clusters}",
      journal = {\apjl},
         year = 2011,
        month = jun,
       volume = {733},
       number = {2},
          eid = {L51},
        pages = {L51},
          doi = {10.1088/2041-8205/733/2/L51},
archivePrefix = {arXiv},
       eprint = {1104.5465},
 primaryClass = {astro-ph.SR},
       adsurl = {https://ui.adsabs.harvard.edu/abs/2011ApJ...733L..51C}
}

@ARTICLE{Casanellas2011a,
       author = {{Casanellas}, Jordi and {Lopes}, Il{\'\i}dio},
        title = "{Towards the use of asteroseismology to investigate the nature of dark matter}",
      journal = {\mnras},
         year = 2011,
        month = jan,
       volume = {410},
       number = {1},
        pages = {535-540},
          doi = {10.1111/j.1365-2966.2010.17463.x},
archivePrefix = {arXiv},
       eprint = {1008.0646},
 primaryClass = {astro-ph.CO},
       adsurl = {https://ui.adsabs.harvard.edu/abs/2011MNRAS.410..535C}
}

@ARTICLE{Freese2008a,
       author = {{Freese}, Katherine and {Bodenheimer}, Peter and {Spolyar}, Douglas and {Gondolo}, Paolo},
        title = "{Stellar Structure of Dark Stars: A First Phase of Stellar Evolution Resulting from Dark Matter Annihilation}",
      journal = {\apjl},
         year = 2008,
        month = oct,
       volume = {685},
       number = {2},
        pages = {L101},
          doi = {10.1086/592685},
archivePrefix = {arXiv},
       eprint = {0806.0617},
 primaryClass = {astro-ph},
       adsurl = {https://ui.adsabs.harvard.edu/abs/2008ApJ...685L.101F}
}

@ARTICLE{Hirano2014,
       author = {{Hirano}, Shingo and {Hosokawa}, Takashi and {Yoshida}, Naoki and {Umeda}, Hideyuki and {Omukai}, Kazuyuki and {Chiaki}, Gen and {Yorke}, Harold W.},
        title = "{One Hundred First Stars: Protostellar Evolution and the Final Masses}",
      journal = {\apj},
         year = 2014,
        month = feb,
       volume = {781},
       number = {2},
          eid = {60},
        pages = {60},
          doi = {10.1088/0004-637X/781/2/60},
archivePrefix = {arXiv},
       eprint = {1308.4456},
 primaryClass = {astro-ph.CO},
       adsurl = {https://ui.adsabs.harvard.edu/abs/2014ApJ...781...60H}
}

@ARTICLE{Hirano2015,
       author = {{Hirano}, S. and {Hosokawa}, T. and {Yoshida}, N. and {Omukai}, K. and {Yorke}, H.~W.},
        title = "{Primordial star formation under the influence of far ultraviolet radiation: 1540 cosmological haloes and the stellar mass distribution}",
      journal = {\mnras},
         year = 2015,
        month = mar,
       volume = {448},
       number = {1},
        pages = {568-587},
          doi = {10.1093/mnras/stv044},
archivePrefix = {arXiv},
       eprint = {1501.01630},
 primaryClass = {astro-ph.GA},
       adsurl = {https://ui.adsabs.harvard.edu/abs/2015MNRAS.448..568H}
}

@ARTICLE{Greif2012,
       author = {{Greif}, Thomas H. and {Bromm}, Volker and {Clark}, Paul C. and {Glover}, Simon C.~O. and {Smith}, Rowan J. and {Klessen}, Ralf S. and {Yoshida}, Naoki and {Springel}, Volker},
        title = "{Formation and evolution of primordial protostellar systems}",
      journal = {\mnras},
         year = 2012,
        month = jul,
       volume = {424},
       number = {1},
        pages = {399-415},
          doi = {10.1111/j.1365-2966.2012.21212.x},
       adsurl = {https://ui.adsabs.harvard.edu/abs/2012MNRAS.424..399G}
}

@ARTICLE{Natarajan2009,
       author = {{Natarajan}, Aravind and {Tan}, Jonathan C. and {O'Shea}, Brian W.},
        title = "{Dark Matter Annihilation and Primordial Star Formation}",
      journal = {\apj},
         year = 2009,
        month = feb,
       volume = {692},
       number = {1},
        pages = {574-583},
          doi = {10.1088/0004-637X/692/1/574},
archivePrefix = {arXiv},
       eprint = {0807.3769},
 primaryClass = {astro-ph},
       adsurl = {https://ui.adsabs.harvard.edu/abs/2009ApJ...692..574N}
}

@ARTICLE{Haemmerle2020,
       author = {{Haemmerl{\'e}}, L. and {Mayer}, L. and {Klessen}, R.~S. and {Hosokawa}, T. and {Madau}, P. and {Bromm}, V.},
        title = "{Formation of the First Stars and Black Holes}",
      journal = {\ssr},
         year = 2020,
        month = apr,
       volume = {216},
       number = {4},
          eid = {48},
        pages = {48},
          doi = {10.1007/s11214-020-00673-y},
archivePrefix = {arXiv},
       eprint = {2003.10533},
 primaryClass = {astro-ph.GA},
       adsurl = {https://ui.adsabs.harvard.edu/abs/2020SSRv..216...48H}
}

@ARTICLE{Stacy2013,
       author = {{Stacy}, Athena and {Greif}, Thomas H. and {Klessen}, Ralf S. and {Bromm}, Volker and {Loeb}, Abraham},
        title = "{Rotation and internal structure of Population III protostars}",
      journal = {\mnras},
         year = 2013,
        month = may,
       volume = {431},
       number = {2},
        pages = {1470-1486},
          doi = {10.1093/mnras/stt264},
archivePrefix = {arXiv},
       eprint = {1209.1439},
 primaryClass = {astro-ph.CO},
       adsurl = {https://ui.adsabs.harvard.edu/abs/2013MNRAS.431.1470S}
}

@ARTICLE{Bromm2009,
       author = {{Bromm}, Volker and {Yoshida}, Naoki and {Hernquist}, Lars and {McKee}, Christopher F.},
        title = "{The formation of the first stars and galaxies}",
      journal = {\nat},
         year = 2009,
        month = may,
       volume = {459},
       number = {7243},
        pages = {49-54},
          doi = {10.1038/nature07990},
archivePrefix = {arXiv},
       eprint = {0905.0929},
 primaryClass = {astro-ph.CO},
       adsurl = {https://ui.adsabs.harvard.edu/abs/2009Natur.459...49B}
}

@ARTICLE{Wise2012,
       author = {{Wise}, John H. and {Turk}, Matthew J. and {Norman}, Michael L. and {Abel}, Tom},
        title = "{The Birth of a Galaxy: Primordial Metal Enrichment and Stellar Populations}",
      journal = {\apj},
         year = 2012,
        month = jan,
       volume = {745},
       number = {1},
          eid = {50},
        pages = {50},
          doi = {10.1088/0004-637X/745/1/50},
archivePrefix = {arXiv},
       eprint = {1011.2632},
 primaryClass = {astro-ph.CO},
       adsurl = {https://ui.adsabs.harvard.edu/abs/2012ApJ...745...50W}
}

@ARTICLE{Scott2009,
       author = {{Scott}, Pat and {Fairbairn}, Malcolm and {Edsj{\"o}}, Joakim},
        title = "{Dark stars at the Galactic Centre - the main sequence}",
      journal = {\mnras},
         year = 2009,
        month = mar,
       volume = {394},
       number = {1},
        pages = {82-104},
          doi = {10.1111/j.1365-2966.2008.14282.x},
archivePrefix = {arXiv},
       eprint = {0809.1871},
 primaryClass = {astro-ph},
       adsurl = {https://ui.adsabs.harvard.edu/abs/2009MNRAS.394...82S}
}

@ARTICLE{Casanellas2011,
       author = {{Casanellas}, Jordi and {Lopes}, Il{\'\i}dio},
        title = "{Signatures of Dark Matter Burning in Nuclear Star Clusters}",
      journal = {\apjl},
         year = 2011,
        month = jun,
       volume = {733},
       number = {2},
          eid = {L51},
        pages = {L51},
          doi = {10.1088/2041-8205/733/2/L51},
archivePrefix = {arXiv},
       eprint = {1104.5465},
 primaryClass = {astro-ph.SR},
       adsurl = {https://ui.adsabs.harvard.edu/abs/2011ApJ...733L..51C}
}

@ARTICLE{Nandal2025c,
       author = {{Nandal}, Devesh and {Topalakis}, Konstantinos and {Tan}, Jonathan C. and {Sergienko}, Vasilisa and {Pauchett}, Ana{\"\i}s and {Petkova}, Maya},
        title = "{The Evolution of Pop III.1 Protostars Powered by Dark Matter Annihilation. I. Fiducial model and first results}",
      journal = {arXiv e-prints},
         year = 2025,
        month = jul,
          eid = {arXiv:2507.00870},
        pages = {arXiv:2507.00870},
          doi = {10.48550/arXiv.2507.00870},
archivePrefix = {arXiv},
       eprint = {2507.00870},
 primaryClass = {astro-ph.SR},
       adsurl = {https://ui.adsabs.harvard.edu/abs/2025arXiv250700870N}
}

@ARTICLE{Nandal2025b,
       author = {{Nandal}, Devesh and {Buldgen}, Ga{\"e}l and {Whalen}, Daniel J. and {Regan}, John and {Woods}, Tyrone E. and {Tan}, Jonathan C.},
        title = "{The Evolution Of Rotating Supermassive Pop III Stars On The Main Sequence}",
      journal = {arXiv e-prints},
         year = 2025,
        month = jun,
          eid = {arXiv:2506.08268},
        pages = {arXiv:2506.08268},
          doi = {10.48550/arXiv.2506.08268},
archivePrefix = {arXiv},
       eprint = {2506.08268},
 primaryClass = {astro-ph.SR},
       adsurl = {https://ui.adsabs.harvard.edu/abs/2025arXiv250608268N}
}

@ARTICLE{Tan2024,
       author = {{Tan}, Jonathan C. and {Singh}, Jasbir and {Cammelli}, Vieri and {Sanati}, Mahsa and {Petkova}, Maya and {Nandal}, Devesh and {Monaco}, Pierluigi},
        title = "{The Origin of Supermassive Black Holes from Pop III.1 Seeds}",
      journal = {arXiv e-prints},
         year = 2024,
        month = dec,
          eid = {arXiv:2412.01828},
        pages = {arXiv:2412.01828},
          doi = {10.48550/arXiv.2412.01828},
archivePrefix = {arXiv},
       eprint = {2412.01828},
 primaryClass = {astro-ph.GA},
       adsurl = {https://ui.adsabs.harvard.edu/abs/2024arXiv241201828T}
}

@ARTICLE{Banik2019,
       author = {{Banik}, Nilanjan and {Tan}, Jonathan C. and {Monaco}, Pierluigi},
        title = "{The formation of supermassive black holes from Population III.1 seeds. I. Cosmic formation histories and clustering properties}",
      journal = {\mnras},
         year = 2019,
        month = mar,
       volume = {483},
       number = {3},
        pages = {3592-3606},
          doi = {10.1093/mnras/sty3298},
archivePrefix = {arXiv},
       eprint = {1608.04421},
 primaryClass = {astro-ph.GA},
       adsurl = {https://ui.adsabs.harvard.edu/abs/2019MNRAS.483.3592B}
}

@ARTICLE{Ilie2025,
       author = {{Ilie}, Cosmin and {Shafaat Mahmud}, Sayed and {Paulin}, Jillian and {Freese}, Katherine},
        title = "{Spectroscopic Supermassive Dark Star candidates}",
      journal = {arXiv e-prints},
         year = 2025,
        month = may,
          eid = {arXiv:2505.06101},
        pages = {arXiv:2505.06101},
          doi = {10.48550/arXiv.2505.06101},
archivePrefix = {arXiv},
       eprint = {2505.06101},
 primaryClass = {astro-ph.CO},
       adsurl = {https://ui.adsabs.harvard.edu/abs/2025arXiv250506101I}
}

@ARTICLE{Maeder2012,
       author = {{Maeder}, Andr{\'e} and {Meynet}, Georges},
        title = "{Rotating massive stars: From first stars to gamma ray bursts}",
      journal = {Reviews of Modern Physics},
         year = 2012,
        month = jan,
       volume = {84},
       number = {1},
        pages = {25-63},
          doi = {10.1103/RevModPhys.84.25},
       adsurl = {https://ui.adsabs.harvard.edu/abs/2012RvMP...84...25M}
}

@ARTICLE{Liu2025,
       author = {{Liu}, Boyuan and {Kessler}, Daniel and {Gessey-Jones}, Thomas and {Dhandha}, Jiten and {Fialkov}, Anastasia and {Sibony}, Yves and {Meynet}, Georges and {Bromm}, Volker and {Barkana}, Rennan},
        title = "{Effects of chemically homogeneous evolution of the first stars on the 21-cm signal and reionization}",
      journal = {\mnras},
         year = 2025,
        month = aug,
       volume = {541},
       number = {4},
        pages = {3113-3133},
          doi = {10.1093/mnras/staf1178},
archivePrefix = {arXiv},
       eprint = {2504.00535},
 primaryClass = {astro-ph.GA},
       adsurl = {https://ui.adsabs.harvard.edu/abs/2025MNRAS.541.3113L}
}

@ARTICLE{Liu2021,
       author = {{Liu}, Boyuan and {Sibony}, Yves and {Meynet}, Georges and {Bromm}, Volker},
        title = "{Stellar winds and metal enrichment from fast-rotating Population III stars}",
      journal = {\mnras},
         year = 2021,
        month = oct,
       volume = {506},
       number = {4},
        pages = {5247-5267},
          doi = {10.1093/mnras/stab2057},
archivePrefix = {arXiv},
       eprint = {2104.10046},
 primaryClass = {astro-ph.GA},
       adsurl = {https://ui.adsabs.harvard.edu/abs/2021MNRAS.506.5247L}
}

@ARTICLE{Maeder-1992,
       author = {{Maeder}, Andre},
        title = "{Stellar yields as a function of initial metallicity and mass limit for black hole formation}",
      journal = {\aap},
         year = 1992,
        month = oct,
       volume = {264},
       number = {1},
        pages = {105-120},
       adsurl = {https://ui.adsabs.harvard.edu/abs/1992A&A...264..105M}
}

@ARTICLE{Patton-2022,
       author = {{Patton}, Rachel A. and {Sukhbold}, Tuguldur and {Eldridge}, J.~J.},
        title = "{Comparing compact object distributions from mass- and presupernova core structure-based prescriptions}",
      journal = {\mnras},
         year = 2022,
        month = mar,
       volume = {511},
       number = {1},
        pages = {903-913},
          doi = {10.1093/mnras/stab3797},
archivePrefix = {arXiv},
       eprint = {2106.05978},
 primaryClass = {astro-ph.HE},
       adsurl = {https://ui.adsabs.harvard.edu/abs/2022MNRAS.511..903P}
}

@ARTICLE{Ilie-2023b,
       author = {{Ilie}, Cosmin and {Freese}, Katherine and {Petric}, Andreea and {Paulin}, Jillian},
        title = "{UHZ1 and the other three most distant quasars observed: possible evidence for Supermassive Dark Stars}",
      journal = {arXiv e-prints},
         year = 2023,
        month = dec,
          eid = {arXiv:2312.13837},
        pages = {arXiv:2312.13837},
          doi = {10.48550/arXiv.2312.13837},
archivePrefix = {arXiv},
       eprint = {2312.13837},
 primaryClass = {astro-ph.GA},
       adsurl = {https://ui.adsabs.harvard.edu/abs/2023arXiv231213837I}
}

@ARTICLE{Chen2020,
       author = {{Chen}, Li-Hsin and {Chen}, Ke-Jung and {Tsai}, Sung-han and {Whalen}, Daniel},
        title = "{How the Population III Initial Mass Function Governs the Properties of the First Galaxies}",
      journal = {arXiv e-prints},
         year = 2020,
        month = oct,
          eid = {arXiv:2010.02212},
        pages = {arXiv:2010.02212},
          doi = {10.48550/arXiv.2010.02212},
archivePrefix = {arXiv},
       eprint = {2010.02212},
 primaryClass = {astro-ph.GA},
       adsurl = {https://ui.adsabs.harvard.edu/abs/2020arXiv201002212C}
}

\end{document}